\documentclass[aps,prl,reprint]{revtex4-2}
\usepackage{amsmath}
\usepackage{amssymb}
\usepackage{graphicx}
\usepackage{mathrsfs}
\usepackage{hyperref}
\hypersetup{colorlinks,linkcolor=blue,citecolor=blue,urlcolor=blue}

\usepackage[caption=false]{subfig}
\usepackage{xcolor}

\begin{document}

\title[Dynamical change under slowly changing conditions: the QKNH Theorem]{Dynamical change under slowly changing conditions:\\
the quantum Kruskal-Neishtadt-Henrard theorem}

\author{Peter Stabel, James R. Anglin}

\address{State reaserch center OPTIMAS and Fachbereich Physik, Technische Universit\"at Kaiserslautern,D-67663 Kaiserslautern, Germany}
\date{\today}

\begin{abstract}
 Adiabatic approximations break down classically when a constant-energy contour splits into separate contours, forcing the system to choose which daughter contour to follow; the choices often represent qualitatively different behavior, so that slowly changing conditions induce a sudden and drastic change in dynamics. The Kruskal-Neishtadt-Henrard theorem relates the probability of each choice to the rates at which the phase space areas enclosed by the different contours are changing. This represents a connection within closed-system mechanics, and without dynamical chaos, between spontaneous change and increase in phase space measure, as required by the Second Law of Thermodynamics. Quantum mechanically, in contrast, dynamical tunneling allows adiabaticity to persist, for very slow parameter change, through a classical splitting of energy contours; the classical and adiabatic limits fail to commute. Here we show that a quantum form of the Kruskal-Neishtadt-Henrard theorem holds nonetheless, due to unitarity.
\end{abstract}
\maketitle

\section{I. Introduction}
When the explicit time dependence of a Hamiltonian is slow compared to the dynamics that the Hamiltonian itself generates, the evolution is usually \emph{adiabatic}. Classically, a single adiabatically evolving degree of freedom follows an energy contour that encloses constant phase space area \cite{Goldstein}. Classical adiabaticity fails at an unstable fixed point, however, where the local dynamical time scale becomes infinite. Unstable fixed points in phase space occur when an energy contour intersects itself; such a self-intersecting energy contour is a \emph{separatrix} dividing phase space into three or more neighboring regions, within which the system dynamics may be qualitatively different. If adiabatic evolution brings a system to a separatrix, the system must choose non-adiabatically which region to enter. The Kruskal-Neishtadt-Henrard (KNH) theorem \cite{Dobrott_Greene,NEISHTADT1975594,henrard_capture_1982} constrains this kind of abrupt change in system evolution due to slow change of the Hamiltonian.

The KNH theorem follows from Liouville's theorem and thus is quite fundamental in classical mechanics. It is potentially useful as the basis of dynamical control strategies that do not require monitoring of a system's state \cite{LiouvilleControl}. As a link within integrable closed-system mechanics between a certain kind of phase space area increase and the probability of spontaneous qualitative change in dynamics, moreover, the KNH theorem may represent the most primitive microscopic limit of the Second Law of Thermodynamics. In this regard its relation to microscopic irreversibility has recently been shown by predicting the probability of small systems to return to their initial configuration after a cyclic parameter sweep (\textit{probabilistic hysteresis})\cite{burkle_how_2019,PhysRevLett.123.114101}. It is hard to accept a classical theorem as a true microscopic precursor to thermodynamics, however: microscopic physics is quantum. Probabilistic hysteresis in quantum systems\cite{PhysRevA.101.042110,PhysRevA.102.052212} has been found numerically to conform to predictions based on the classical KNH theorem, but only when the initial state is a sufficiently wide ensemble of energy levels and only for sweeps that are not too slow. For infinitely slow cyclic parameter sweeps, the quantum adiabatic theorem\cite{BF,kato,Simon} forbids hysteresis and breaks correspondence with the classical KNH theorem. The extension to quantum mechanics of the KNH theorem, with its microscopic resemblance to thermodynamics, is therefore non-trivial.
\begin{figure}
	\centering
	\includegraphics[width=0.48\textwidth]{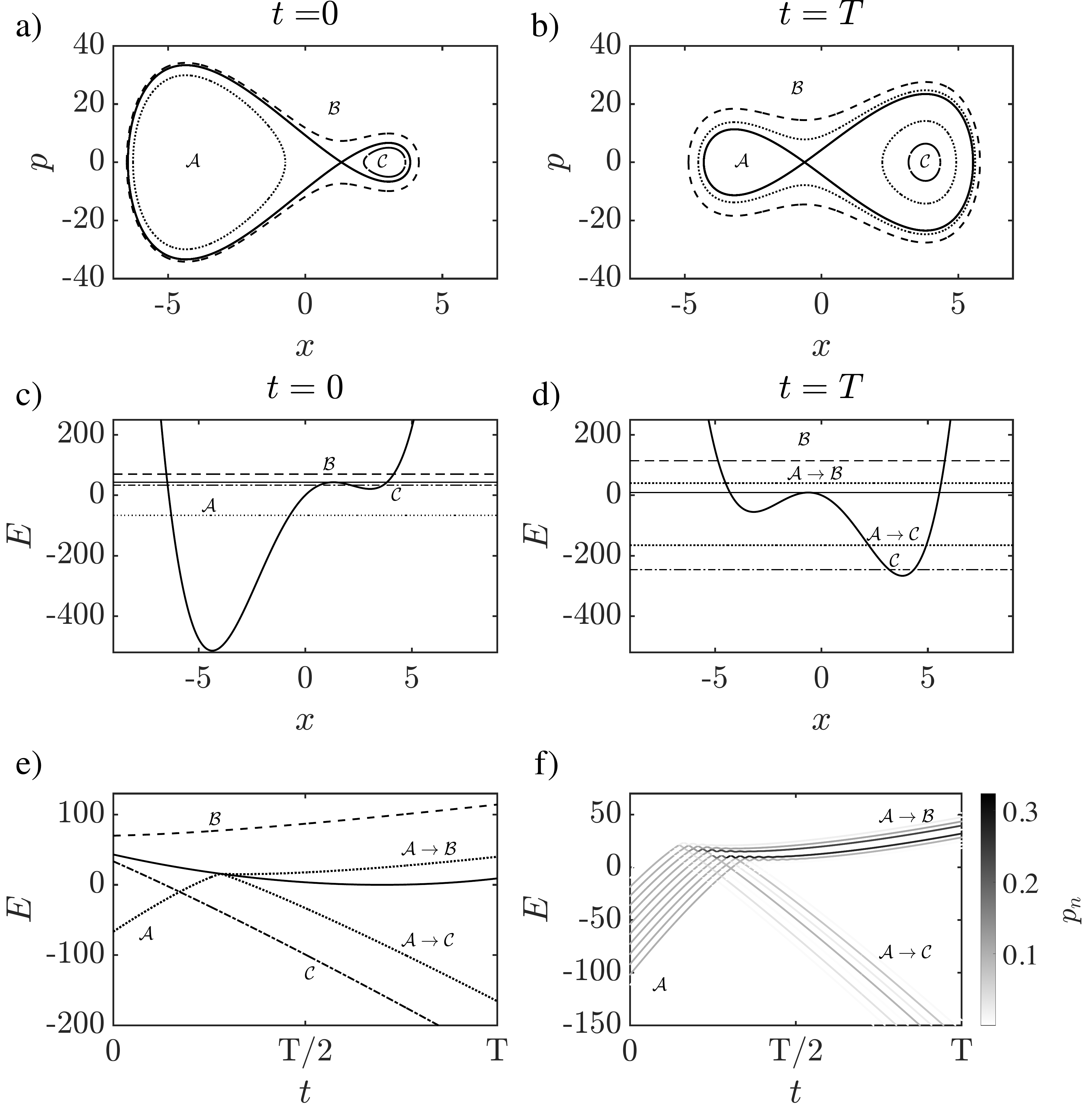}
	\caption{A slowly time-dependent separatrix divides $(x,p)$ phase space into three regions $\mathcal{A}$, $\mathcal{B}$, and $\mathcal{C}$, shown at $t=0$ (a) and at a later time $T$ (b). This could represent a time-dependent double well potential (c, d), with the three phase space regions corresponding to trajectories within either well ($\mathcal{A}$ and $\mathcal{C}$) or with enough energy to cross the barrier ($\mathcal{B}$). Panel (e) shows energy versus continuous time for the orbits shown at the two snapshot times $0,T$ in (a-d); orbits which hit the separatrix are distributed post-adiabatically among orbits above and below the separatrix in energy. The analogous quantum process for level probabilities (f) is our subject. The double well potential used for  this figure is of the form $V(x,t) = \alpha x^4 - \beta(\lambda(t))x^2 + \gamma(\lambda(t))x$; we keep $\alpha$ constant and sweep $\lambda$ linearly in time.}
	\label{FigDblWell}
\end{figure}
Since the KNH theorem concerns energy contours in phase space, its extension to quantum mechanics must begin in the semi-classical limit where the Wentzel-Kramers-Brillouin-Jeffreys (WKBJ) approximation relates energy contours in phase space to quantum energy eigenstates. The quantum KNH theorem cannot be deduced just from quantum-classical correspondence, however, because this correspondence breaks down at unstable fixed points, and because the adiabatic and semi-classical limits in quantum mechanics do not commute \cite{Berry_1984,Wu_Liu}. Here we show how the KNH theorem extends to quantum post-adiabatic dynamics, starting from WKBJ semiclassical theory.

\section{Adiabatic change and post-adiabatic choice}
The general scenario is illustrated in Fig.~\ref{FigDblWell}. As an example of a Hamiltonian with a separatrix, we consider a double well potential $V(x,\lambda)$ which depends on a parameter $\lambda(t)$ that increases with time $t$ slowly and monotonically. Phase space orbits inside each lobe of the separatrix, shown in dots and long dashes, have energies below the height $V_b(\lambda)$ of the central barrier, and are therefore confined (classically) in one well or the other. A phase space orbit outside the separatrix, shown in short dashes, has energy above the barrier, so that the system traverses both wells. If the potential changes slowly, the adiabatic theorem states that the system's energy changes so as to hold the phase space area inside its orbit constant. 

The adiabatic theorem does not apply to the separatrix; its energy simply equals the instantaneous barrier height $V_b(\lambda)$, and its enclosed area may change as the potential changes. Very near the separatrix, moreover, the adiabatic theorem does not apply to the system, either, because the system moves too slowly near the unstable fixed point. Evolution under the time-dependent Hamiltonian can thus bring the system \emph{across} the separatrix, even though the separatrix is an energy contour. 

A separatrix lobe may expand into and absorb adiabatic orbits of the system; conversely it can contract and squeeze orbits out, as in the case of the dotted orbits in Fig.~\ref{FigDblWell}. A system which crosses a slowly changing separatrix may therefore have to choose between different kinds of new orbits. If the barrier rises, for example, a system that is initially orbiting above the barrier may be captured into one well---or the other. If one well is becoming narrower or shallower, a system which is initially trapped in it may either be tipped into the other well, or excited above the barrier. An example of this latter scenario, with both outcomes occurring for different initial states that all have the same energy, is shown in Fig.~\ref{FigDblWell}. Realizations of these basic dynamical processes range from satellite capture to chemical reactions. 

\subsection{The KNH theorem}
Because the crux of the separatrix is an \emph{unstable} fixed point of the instantaneous Hamiltonian, the system's fate after crossing a separatrix depends sensitively on its initial conditions, as well as on exactly how (and how slowly) the potential changes. If the rate at which the Hamiltonian changes is much slower than the rate of exponential approach/departure at the unstable fixed point, however, then there is a simple rule governing the \emph{fractions} of initial states which evolve post-adiabatically into each of the three phase space regions that the separatrix defines (two lobes and the exterior).

The KNH theorem states firstly that orbits can only leave an adiabatic region which is shrinking in phase space area, and can only move into an adiabatic region which is growing in area. The theorem further states that if there is more than one growing adiabatic region then the fractions of orbits which enter each growing region, from a shrinking region, are proportional to the rates at which the growing areas grow. If region $\mathcal{A}$ is the only shrinking region and region $\mathcal{C}$ is growing, for example, 
\begin{equation}
    P_{\mathcal{A}\to \mathcal{C}}=-\frac{\frac{d}{d\lambda} S_\mathcal{C}\big(V_b(\lambda),\lambda\big)}{\frac{d}{d\lambda} S_\mathcal{A}\big(V_b(\lambda),\lambda\big)}\;,
\end{equation}
where the barrier height $V_b$ is also the classical separatrix energy; the Hamiltonian for a particle of mass $\mu$ in the potential is $H(x,p,\lambda)=p^2/(2\mu) + V(x,\lambda)$; $\mathcal{A}$ is the donor region (such as a shrinking lobe of the separatrix) shown in Fig.\ref{FigDblWell}a) and b); $\mathcal{C}$ is one recipient region (such as the growing other lobe in the Figure); and $S_\mathcal{A,C}(E,\lambda)$ is the area enclosed in either region by the contour $H(x,p,\lambda)=E$.

As soon as it is stated the KNH theorem may seem to be an obvious consequence of the fact that Hamiltonian evolution is an incompressible flow in phase space, according to Liouville's theorem. Issues such as the time at which the area growth rates are to be calculated, and the canonical coordinates which should be used, are subtle, however, and a rigorous proof has only been provided quite recently \cite{Neishtadt_2017}. Obvious or not, the classical KNH theorem provides a direct dynamical connection between phase space area growth and spontaneous qualitative change, reminiscent of the Second Law of thermodynamics, even though the enclosed areas to which the KNH theorem refers are \emph{not} ergodically explored by the system, and no assumptions about equilibration are made.

\subsection{Non-trivial quantum correspondence}
Away from separatrices, where the classical orbit period $\partial_{E}S$ remains finite, a single quantum degree of freedom in the semi-classical limit obeys Bohr-Sommerfeld energy quantization, which implies that the spacing between successive energy levels is $\pi\hbar/\partial_{E}S$. The conditions for quantum and classical adiabaticity therefore typically coincide away from separatrices, both being satisfied when $\lambda(t)$ changes slowly on the time scale of $\partial_{E}S$. Near a separatrix, however, this quantum-classical correspondence of adiabaticity breaks down.

Although classically the orbital period diverges at the separatrix, and so for any finitely slow $\lambda(t)$ adiabaticity must fail within a finite neighborhood of the separatrix, quantum mechanical energy levels generally do \emph{not} become degenerate at the barrier height $E=V_b$. Although the partial derivatives $\partial_E S_\mathcal{A,C}(E,\lambda)$ generically diverge logarithmically as $E\to V_b$, correctly supplementing the WKBJ semi-classical theory with connection formulas through the classical turning points and allowing for tunneling \cite{CHILD1974280} leads to the modified Bohr-Sommerfeld quantization condition for $E<V_b$
\begin{equation}\label{GenBS}
\cos\Big(\frac{\tilde{S}_\mathcal{A}-\tilde{S}_\mathcal{C}}{\hbar}\Big) \stackrel{!}{=}-\sqrt{1 +e^{-2T_b/\hbar}}\cos\Big(\frac{\tilde{S}_\mathcal{A}+\tilde{S}_\mathcal{C}}{\hbar}\Big)\;.\end{equation}
Here\begin{equation}\label{Tb}
T_b(E,\lambda)=\int_{x_1(E,\lambda)}^{x_2(E,\lambda)}\!dx\,\sqrt{2\mu[V(x,\lambda)-E]}\end{equation}
is the non-classical action associated with tunneling through the potential barrier between classical turning points $x_{1,2}(E,\lambda)$, and $\tilde{S}_\mathcal{A,C}$ are \emph{quantum-corrected} versions of the classical areas $S_\mathcal{A,C}$ \cite{CHILD1974280} (see \ref{app:bs}). In the semi-classical limit of action scales much larger than $\hbar$, the quantum correction term in $\tilde{S}$ is generally negligible except for energies close to the barrier height, but it is enough to keep  $\partial_E\tilde{S}_\mathcal{A,C}$ \emph{finite} at $E=V_b$, so WKBJ energy spacings do \emph{not} all become small near $E=V_b$ and there is no general breakdown of adiabaticity in the WKBJ limit. This is an example of the general fact that the adiabatic and classical limits do not commute \cite{Berry_1984,Wu_Liu}. 
\begin{figure}
	\centering
	\includegraphics[width = 0.45\textwidth]{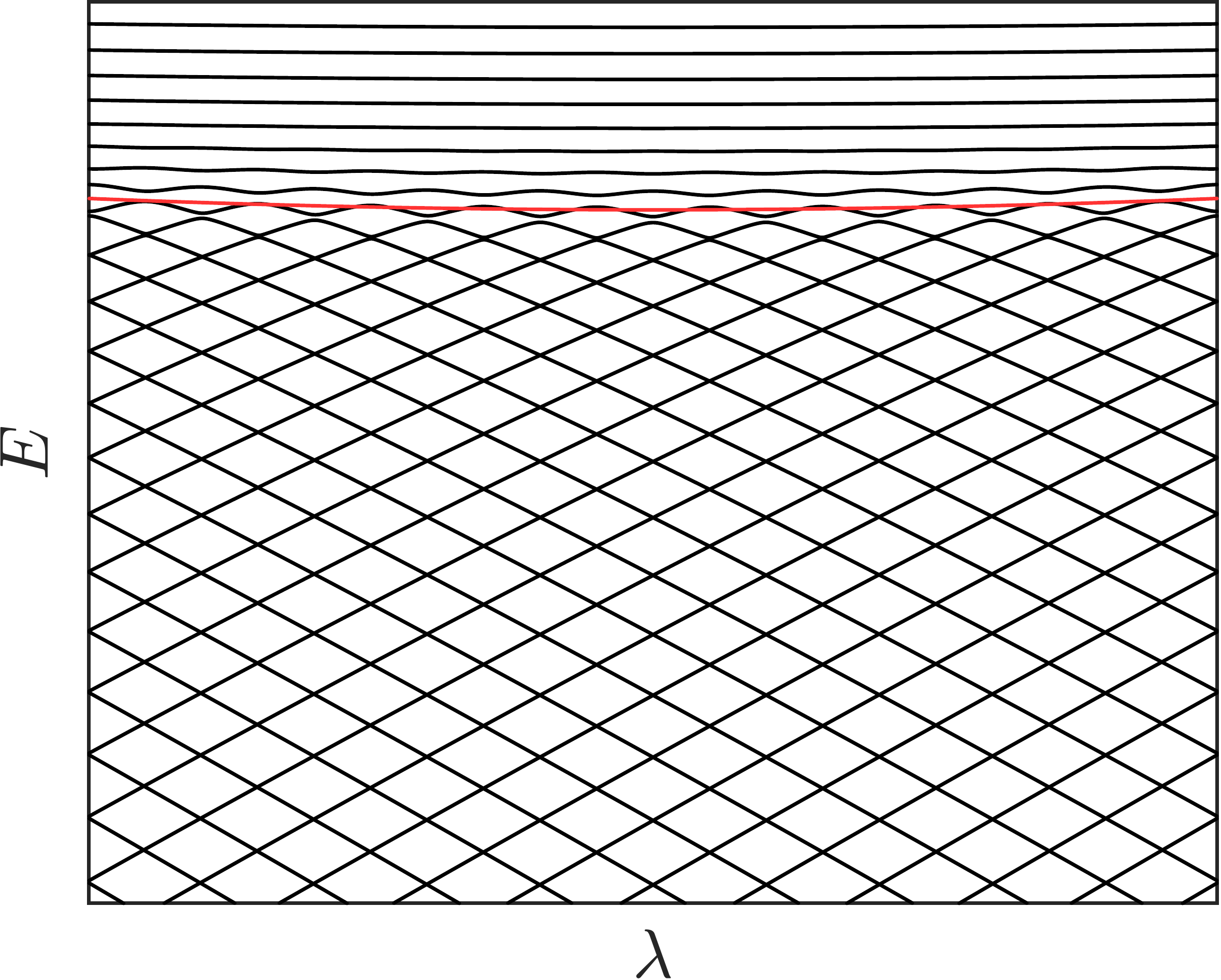}
	\caption{Numerical eigenspectrum of a $\lambda$-dependent quantum double well, in a portion of the $(E,\lambda)$ plane. The quantum levels (black curves) exhibit a clear change in character around the classical separatrix (red curve), but they do not become degenerate there. Instead they have what appears to be a lattice of crossings below the separatrix energy; in fact these crossings are all avoided, with gaps too small to be visible here. Our analysis concerns these kinds of avoided crossing lattices, in the semi-classical limit.}
	\label{QsepFig}
\end{figure}
An example to show how numerically exact quantum energy levels $E_n(\lambda)$ do conform to this semi-classical picture is shown in Fig.~\ref{QsepFig}. Since the KNH theorem explicitly concerns phase space areas, we can expect that its quantum form is defined in the classical correspondence limit, but can the KNH theorem emerge from quantum mechanics at all, if quantum adiabaticity does not actually fail at barrier tops?

\section{Avoided crossings in the $(E,\lambda)$ plane}
The KNH theorem does emerge from quantum mechanics, because even though adiabaticity does not fail near $E=V_b$ for all $\lambda$ quantum mechanically as it does classically, it does fail at a certain discrete set of special $\lambda$ values, for energies less than $V_b$ by some finite amount, whenever $e^{-2T_b/\hbar}\ll 1$. 

\subsection{The lattice of avoided crossings}
To see why this is, note that to zeroth order in $e^{-2T_b/\hbar}$ the WKBJ condition (\ref{GenBS}) is satisfied by \emph{either} $\tilde{S}_\mathcal{A} = (\tilde{m}+1/2)\pi\hbar$ \emph{or} $\tilde{S}_\mathcal{C} = (\tilde{n}+1/2)\pi\hbar$ for integers $\tilde{m},\tilde{n}$. To avoid having to use large integers to label the high $\tilde{m},\tilde{n}$ states in which we will be interested in this paper, we define $\tilde{m}=m+m_0$ and $\tilde{n}=n+n_0$ for some integer shifts $m_0,n_0$, and use integers $m,n$ as our quantum numbers, which may be negative.
Since $\tilde{S}_\mathcal{A,C}$ are in general different functions of $E$ and $\lambda$, there are two sets of energy levels $E^\mathcal{A}_{m}(\lambda)$, $E^\mathcal{C}_{n}(\lambda)$. Successive levels within each set are not degenerate, $E^\mathcal{A}_{m+1}>E^\mathcal{A}_m$ and $E^\mathcal{C}_{n+1}>E^\mathcal{C}_n$, but only if perfect symmetry of $V(x,\lambda)$ is maintained as $\lambda(t)$ changes will the two sets of levels shift with $\lambda$ in parallel. To zeroth order in $e^{-2T_b/\hbar}$ there will generically be a lattice of almost-crossings, $E^\mathcal{A}_m(\lambda_{mn})- E^\mathcal{C}_n(\lambda_{mn})=\mathcal{O}(e^{-2T_b/\hbar})$, at a discrete set of parameter values $\lambda_{mn}$. In the semi-classical limit, therefore, the discrete quantum energy spectrum in the range $E<V_b$ forms a lattice with $\mathcal{O}(\hbar)$ unit cell size. 

\begin{figure}
	\centering
	\includegraphics[width=0.48\textwidth]{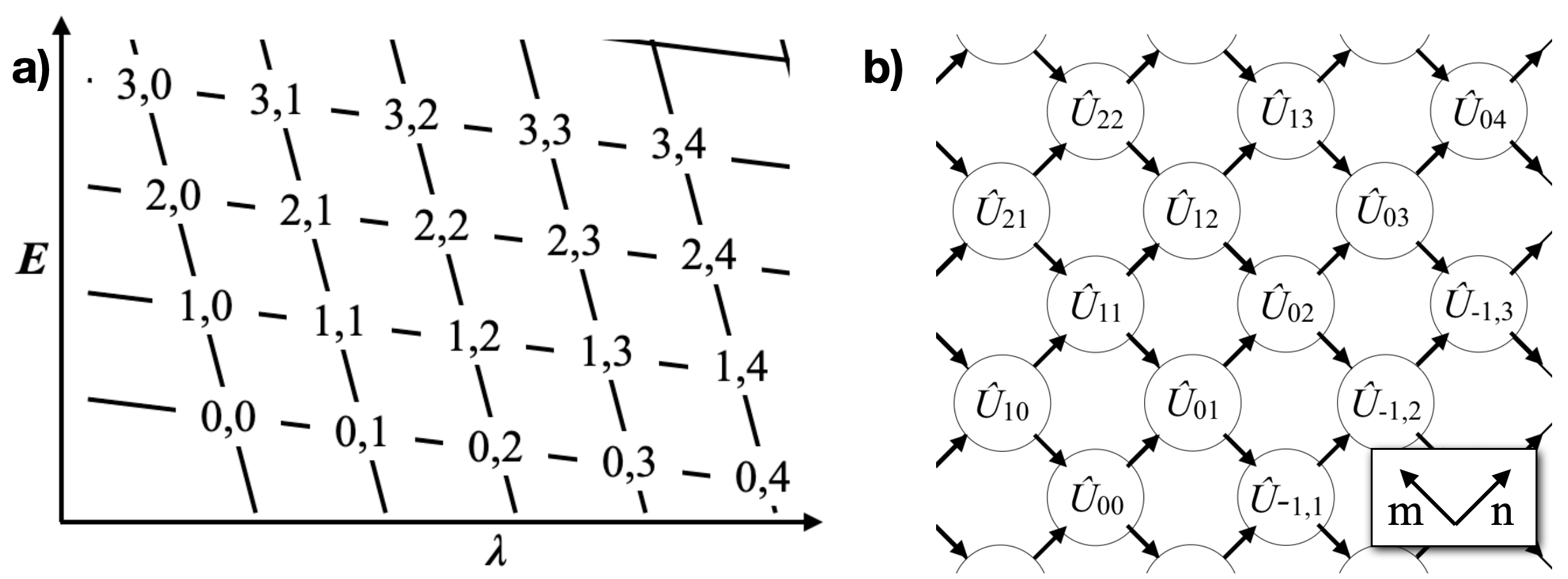}
	\caption{a) Bohr-Sommerfeld energy eigenvalues in the $(E,\lambda)$ plane form a lattice which in the semi-classical limit is locally regular, like the exact lattice seen in the lower part of Fig.~\ref{QsepFig}. In general the parallel $E^\mathcal{A}_m$ lines and the parallel $E^\mathcal{C}_n$ lines may have arbitrary slopes. The $(m,n)$  Bohr-Sommerfeld crossings are actually narrowly avoided because of tunneling. b) The $(E,\lambda)$ plane may thus be represented with continuous $(m,n)$ as skewed and tilted coordinates. Since the time evolution is adiabatic between the avoided crossings at the integer $(m,n)$, but may be non-adiabatic through each avoided crossing, the evolution reduces to a discrete feed-forward network of pairwise unitary transformations $\hat{U}_{mn}$ that each act only in a two-dimensional subspace of crossing Bohr-Sommerfeld levels.}
	\label{LatticeFig}
\end{figure}

The lattice is locally regular in the sense that its curvature and non-uniformity only become non-negligible over $\mathcal{O}(\hbar^{-1})\gg 1$ lattice spacings. If $(E_{00},\lambda_{00})$ is the location in the $(E,\lambda)$ plane of one of these near-crossings, then the lattice of nearby near-crossings is given (see \ref{app:lattice_avoided_crossings}) by
\begin{align}\label{Elam2}
    E_{mn}-E_{00} &= \pi\hbar \frac{m\partial_\lambda \tilde{S}_\mathcal{C} - n\partial_\lambda\tilde{S}_\mathcal{A}}{[\tilde{S}_\mathcal{A},\tilde{S}_\mathcal{C}]} +\mathcal{O}(m\hbar,n\hbar)^2\nonumber\\
    \lambda_{mn}-\lambda_{00} &=\pi\hbar\frac{n\partial_E \tilde{S}_\mathcal{A} - m\partial_E\tilde{S}_\mathcal{C}}{[\tilde{S}_\mathcal{A},\tilde{S}_\mathcal{C}]} +\mathcal{O}(m\hbar,n\hbar)^2\;.
\end{align}
when we introduce the Poisson-like bracket
\begin{equation}\label{poisson}
[F,G]:= \frac{\partial F}{\partial E}\frac{\partial G}{\partial\lambda}-\frac{\partial G}{\partial E}\frac{\partial F}{\partial\lambda}\;.
\end{equation}
In between these near-crossings the semi-classical energy levels follow curves in the $(E,\lambda)$ plane that can be well approximated as straight lines over many lattice cells, as illustrated in  Figs.~\ref{QsepFig} and \ref{LatticeFig}a).

As we review in \ref{app:hamiltonian}, when the tunneling factor $e^{-2T_b/\hbar}$ is \emph{not} neglected then in fact $E^\mathcal{A}_m(\lambda_{mn})- E^\mathcal{C}_n(\lambda_{mn})\not= 0$: the crossings are \emph{avoided} due to quantum tunneling. The two-state Hamiltonian for each two nearly-crossing levels, for $\lambda$ near the zeroth-order crossing point $\lambda_{mn}$, is actually
\begin{align}\label{LZmodel}
    \hat{H}(t)=&\,\bar{E} + \hbar\nu^2 (t-t_{mn}) \frac{\hat{\sigma}_z}{2}+ \hbar\gamma\frac{\hat{\sigma}_x}{2}\\
  \nu^2= &\,
    \frac{1}{\hbar}\frac{d\lambda}{dt}\left|\frac{dE^\mathcal{A}_m}{d\lambda}-\frac{dE^\mathcal{C}_n}{d\lambda}\right|\quad
    \gamma=\frac{e^{-T_b/\hbar}}{\sqrt{\partial_E S_\mathcal{A}\partial_E S_\mathcal{C}}}\nonumber\\
    \hat{\sigma}_z = & \,\mathrm{sgn}\!\left(\frac{dE^\mathcal{A}_m}{d\lambda}-\frac{dE^\mathcal{C}_n}{d\lambda}\right)\left(|E^\mathcal{A}_m\rangle\langle E^\mathcal{A}_m| - |E^\mathcal{C}_n\rangle\langle E^\mathcal{C}_n|\right)\nonumber\\
    \hat{\sigma}_x =&\, |E^\mathcal{A}_m\rangle\langle E^\mathcal{C}_n| + |E^\mathcal{C}_n\rangle\langle E^\mathcal{A}_m|\nonumber
\end{align}
where $\bar{E}=E_{mn}+(\lambda-\lambda_{mn})d(E^\mathcal{A}_m+E^\mathcal{C}_n)/d\lambda$ and $E_{mn}$ is the energy at which $E^\mathcal{A}_m$ and $E^\mathcal{C}_n$ cross at zeroth order, and $\lambda(t_{mn})=\lambda_{mn}$ defines the time $t_{mn}$ at which this crossing is reached. All functions of $\lambda$ and $E$ in (\ref{LZmodel}) are to be evaluated at $(E,\lambda)$ = $(E_{mn},\lambda_{mn})$, and $d{\lambda}/dt$ is to be evaluated at $t_{mn}$. The two instantaneous eigenvalues of this $\hat{H}$ are easily computed as
\begin{equation}
    E_\pm = \bar{E} \pm \frac{\hbar}{2}\sqrt{\nu^4(t-t_{mn})^2+\gamma^2}\;, 
\end{equation}
which are separated by a minimum gap of $\hbar\gamma$, but approach $E^\mathcal{A}_m$ and $E^\mathcal{C}_n$ for large $|t-t_{mn}|$, with the continuous eigenvalue $E_\pm(t)$ that coincides with $E^\mathcal{A}_m(\lambda)$ at large negative $t-t_{mn}$ becoming equal to $E^\mathcal{C}_n(\lambda)$ at large positive $t-t_{mn}$, and \emph{vice versa}.

Where the energy gap between two instantaneous energy eigenstates becomes small, the quantum adiabatic approximation may break down, depending on how rapidly the Hamiltonian depends on time. The breakdown of adiabaticity remains simple, however, inasmuch as it only concerns energy levels that are becoming nearly degenerate. Quantum time evolution through the intervals around each $t_{mn}$ includes a non-trivial unitary evolution $\hat{U}_{mn}$ within each two-dimensional subspace of crossing levels, which can be represented in general as
\begin{align}\label{Umn}
    &\hat{U}_{mn} = e^{-ia} e^{-ib\hat{\sigma}_z}\hat{W}e^{-ic\hat{\sigma}_z}\\
    &\hat{W}=\big(|E^\mathcal{A}_m\rangle\;,\;|E^\mathcal{C}_n\rangle\big)\left(\begin{matrix}\sqrt{P_{mn}} & \sqrt{1-P_{mn}}\\ -\sqrt{1-P_{mn}} & \sqrt{P_{mn}}\end{matrix}\right)\left(\begin{matrix}\langle E^\mathcal{A}_m|\\ \langle E^\mathcal{C}_n|\end{matrix}\right)\nonumber 
\end{align}
where $\hat{\sigma}_z$ is as in (\ref{LZmodel}) and the three angles $a,b,c$ as well as the operator $\hat{W}$ can be different for each $m,n$ (subscripts $_{mn}$ on $a,b,c$ and $\hat{W}$ are left implicit to keep the formulas legible).

For two-state avoided crossings like (\ref{LZmodel}), the non-perturbative Landau-Zener formula \cite{LZ} yields  (\ref{Umn}) with $P_{mn}=e^{-2\pi\gamma^2/\nu^2}=: P(E_{mn},\lambda_{mn})$. In our particular case (\ref{LZmodel}), this probability can be expressed (see \ref{app:probability_lattice}) in the form
\begin{equation}\label{PElam}
    P(E,\lambda) = \exp\left[-\frac{\pi\hbar\, \exp\big(-2T_b(E,\lambda)/\hbar\big)}{\dot{\lambda}\Big|[\tilde{S}_\mathcal{A},\tilde{S}_\mathcal{C}]\Big|}\right]
    \;,
\end{equation}

$P_{mn}$ in (\ref{Umn}) represents the probability for a \emph{diabatic} evolution through the avoided crossing, in which the system emerges on the same energy line along which it approached the vertex ($|E^\mathcal{A}_m\rangle\to |E^\mathcal{A}_m\rangle$ and $|E^\mathcal{C}_n\rangle\to |E^\mathcal{C}_n\rangle$), while $1-P_{mn}$ is the probability for adiabatic quantum evolution through the avoided crossing, following the same $\lambda$-dependent eigenstate of $\hat{H}$ from (\ref{LZmodel}) as it continuously rotates from $|E^\mathcal{A}_m\rangle$ to $|E^\mathcal{C}_n\rangle$ or \emph{vice versa}. 

In fact $\hat{U}_{mn}$ is not simply a classical random choice between two outcomes, but describes evolution into coherent superpositions of the adiabatic states, with amplitude phases given by $a,b,c$. Evolution under the $\hat{H}$ of (\ref{LZmodel}) for arbitrarily long times implies particular time-dependent forms of $a,b,c$, but (\ref{LZmodel}) only holds for each $m,n$ while $\lambda(t)$ is close to $\lambda_{mn}$; after each such interval there is some general adiabatic evolution, for the particular $E^\mathcal{A}_m$ and $E^\mathcal{C}_n$ levels, in the particular $V(x,\lambda)$ potential, until the next avoided crossing is approached. Whatever this general adiabatic evolution is, however, it can be absorbed into the $a$, $b$ and $c$ phases of each $\hat{U}_{mn}$. The entire quantum evolution, adiabatic except possibly at avoided crossings, can thus be represented without loss of generality as a feed-forward linear network of unitary transformations at $(m,n)$ nodes, as illustrated in Fig.~\ref{LatticeFig}b). The quantum KNH theorem concerns this unitary transition network, within which we can identify a quantum analog to the classical separatrix.

\subsection{The quantum separatrix}
A classical separatrix is usually considered as a curve in phase space, but in an adiabatic problem with a time-dependent parameter $\lambda(t)$, the classical separatrix can also be represented as the curve $E=V_b(\lambda)$ in the $(E,\lambda)$ plane, as marked in Fig.~1e). Following the latter concept of a separatrix, we define the \emph{quantum separatrix} $E_s(\lambda)$ to be the curve in the $(E,\lambda)$ plane on which 
\begin{equation}\label{Psep}
P(E_s,\lambda)= 1/e. 
\end{equation}
The reason for this definition appears if we examine the avoided crossings in the neighborhood of any point on the quantum separatrix. As the lattice origin point $(m,n)=(0,0)$ we select an arbitrary crossing which is closer than any of its neighbors to the separatrix. We then use (\ref{Elam2}) in $P_{mn}=P({E}_{mn},\lambda_{mn})$ to conclude for $|m,n|\ll\mathcal{O}(\hbar^{-1})$ 
\begin{align}\label{Pmn}
P_{mn}=&\exp[-Ze^{mX}e^{nY}]\nonumber\\
Z= &\,\frac{2\pi\hbar\, \exp[-2T_b(E_{00},\lambda_{00})/\hbar]}{\dot{\lambda}\Big|[\tilde{S}_\mathcal{A},\tilde{S}_\mathcal{C}]\Big|}\nonumber\\
X=&\, 2\frac{[T_b,\tilde{S}_\mathcal{C}]}{[\tilde{S}_\mathcal{A},\tilde{S}_\mathcal{C}]} \qquad
Y= 2\frac{[T_b,\tilde{S}_\mathcal{A}]}{[\tilde{S}_\mathcal{C},\tilde{S}_\mathcal{A}]}\;,
\end{align}
up to correction factors $\times[1+\mathcal{O}(\hbar)]$ in the exponent of $P_{mn}$. The formulas for $X$ and $Y$ in (\ref{Pmn}) are thus to be evaluated at $(E,\lambda)=(E_{00},\lambda_{00})$. Importantly, $X$ and $Y$ are of order $\hbar^0$. 

Equation (\ref{Pmn}) is the basis for all the main results of this paper; see \ref{app:probability_lattice} for its detailed derivation. It holds in the semi-classical limit, where we neglect corrections of $\mathcal{O}(\hbar)$, and it holds with constant $X,Y,Z$ over the range $|m|,|n|\ll\mathcal{O}(\hbar^{-1})$ over which the lattice of avoided crossings can be approximated as regular. Over larger $\mathcal{O}(\hbar^{-1})$ ranges of $m,n$, $X$, $Y$, and $Z$ can be considered as slowly varying; they are local characteristics of the avoided crossing lattice. Here we will consider only ensembles of initial states within a narrow enough energy range for any $m,n$-dependence of $X$, $Y$, and $Z$ to be neglected. 

For avoided crossings with ${E}_{mn} < E_s(\lambda_{mn})$ (below the separatrix), the double-exponential function $P_{mn}\to 1$ within a distance from the separatrix in $(m,n)$ lattice units of order $\hbar^0$, while for ${E}_{mn}>E_s(\lambda_{mn})$, $P_{mn}\to 0$ within a similar distance. The energy band between these limits, within which transitions are neither very adiabatic nor very diabatic ($-\ln(P_{mn})=\mathcal{O}(1)$), is the \emph{quantum separatrix zone}. The precise width (number of energy levels $D$) of the separatrix zone thus depends on how small a $P_{mn}$, or a $1-P_{mn}$, we are prepared to ignore, but the double-exponential form of $P_{mn}$ ensures that arbitrarily small $P_{mn}$ and $1-P_{mn}$ are reached within a number of lattice spacings $D$ that is not large unless $X$ and $Y$ are both anomalously small. Although the width of the separatrix zone is thus not quite precisely defined, its width in energy is $D\times\mathcal{O}(\hbar)$, very narrow in classical terms. Unlike the classical separatrix, the location of the quantum separatrix in the $E,\lambda$ plane depends on the sweep rate $d{\lambda}/dt$. For very slow $\lambda(t)$, $E_s$ can fall well below the classical separatrix energy $V_b$. And $E_s(\lambda)$ can also change with $\lambda(t)$ because $\dot{\lambda}$ changes, as well as because $V(x,\lambda)$ changes.

Wherever the separatrix zone is, and however wide it is, below it the system passes through every avoided crossing \emph{diabatically}, in the sense that the Landau-Zener probability of an adiabatic transition is negligible. Although quantum mechanically these are non-adiabatic transitions, the result is similar to adiabatic classical evolution, with the system remaining always in either left-well eigenstates with $E^\mathcal{A}_m$ or right-well eigenstates with $E^\mathcal{C}_n$. Above the separatrix zone, in contrast, the crossings are all instead \emph{adiabatic}; the system zig-zags through the $(E,\lambda)$ lattice, alternating between left-well and right-well states, by tunneling back and forth through the barrier at every crossing. The quantum separatrix is thus also, like the classical one, a division between three qualitatively different kinds of dynamics: localized in either left or right well, below the separatrix, or passing through both wells, above it. We will therefore retain our $\mathcal{A,B,C}$ labels of the three classical phase space regions, and use them henceforth to refer to these three dynamically distinct $\lambda$-dependent subspaces of the quantum energy spectrum, along with the separatrix zone $\mathcal{S}$ as a fourth subspace.

The concept of a separatrix between qualitatively different forms of dynamics thus does extend from classical mechanics into quantum mechanics, along with the breakdown of simple adiabatic behavior within a narrow zone around the separatrix. This extension of the separatrix concept survives in spite of the fact that quantum tunneling preserves adiabaticity at the classical separatrix; indeed we might say that the quantum separatrix exists precisely because of tunneling, since it depends on the narrow avoidance of level crossing that tunneling creates. Although (\ref{Pmn}) holds in the limit $\hbar\to 0$, and in this sense represents behavior as close to classical as quantum evolution near a separatrix can be, it still consists of probabilities for superpositions of discrete energy levels that coherently mix due to tunneling. This remarkably non-classical form of classical limit illustrates the subtlety of combining adiabaticity, instability, and quantum-classical correspondence; and yet we will see how behavior similar to classical emerges from it.

\subsection{Growth conditions} 
While the dimension $D$ of the separatrix zone subspace $\mathcal{S}$ is by definition essentially constant over many $(m,n)$ lattice cells, the sizes of the three $\mathcal{A,B,C}$ subspaces are generally changing with $\lambda$. In the $(m,n)$ plane, the separatrix runs parallel to the vector $(-Y,X)$, according to (\ref{Pmn}), while $\lambda$ is given in terms of $(m,n)$ by (\ref{Elam2}). With a bit of two-variable calculus (see \ref{app:growthrates} ) we obtain for the average rates of change of the dimensionalities $D_\mathcal{A,B,C}$ of the respective subspaces
\begin{align}\label{Nrates}
    \frac{dD_\mathcal{A}}{d\lambda} &= -\Gamma Y\nonumber\\
    \frac{dD_\mathcal{B}}{d\lambda} &= \Gamma (Y-X)\nonumber\\
    \frac{dD_\mathcal{C}}{d\lambda} &= \Gamma X\nonumber\\
    \mathrm{for}\quad\Gamma &=\frac{[\tilde{S}_\mathcal{A},\tilde{S}_\mathcal{C}]}{\pi\hbar(X\partial_E\tilde{S}_\mathcal{A}+Y\partial_E\tilde{S}_\mathcal{C})}\;.
\end{align}
These change rates necessarily sum to zero since the size $D$ of the separatrix zone, and that of the whole Hilbert space, are independent of $\lambda$. Given Bohr-Sommerfeld quantization, these growth rates in subspace dimensionality correspond directly (with a factor of $\pi\hbar$) to the growth rates of (quantum-corrected) phase space areas. The quantum KNH theorem will therefore express probabilities for transitions between different subspaces in terms of ratios among the parameters $X$, $Y$, and $Y-X$.

Analogously to the classical case, the first part of the quantum KNH theorem constrains when transitions between subspaces can be possible at all. Suppose, for example, that subspace $\mathcal{A}$ is growing, $dD_\mathcal{A}/d\lambda \propto -Y > 0$. This means that $E_S-E^\mathcal{A}_m$ is growing for any fixed $m$ and new, higher-$m$ $|E^\mathcal{A}_m\rangle$ are entering the $\mathcal{A}$ subspace because their transitions at crossings are becoming sufficiently diabatic. For all the $|E^\mathcal{A}_m\rangle$ that were already in the $\mathcal{A}$ subspace ($E^\mathcal{A}_m < E_S$), the crossings are only becoming more perfectly diabatic as $\lambda$ increases. So if the system is in any growing subspace, it cannot leave the subspace. And by repeating this same argument with time reversed we conclude that the system can never enter a subspace which is shrinking. So the first KNH rule carries over to quantum mechanics in the WKBJ limit simply with $S\to\tilde{S}$.

If a subspace is shrinking, conversely, then the system can be forced to exit that subspace, if it occupies one of the upper (for $\mathcal{A}$ and $\mathcal{C}$) or lower (for $\mathcal{B}$) energy levels in the subspace which is approaching the separatrix. If $dD_\mathcal{A}/d\lambda \propto -Y < 0$, for instance, this means that as $\lambda$ increases the transitions at crossings are steadily becoming less perfectly diabatic, until the amplitude for an adiabatic transition can no longer be ignored: one after another the uppermost $|E^\mathcal{A}_m\rangle$ levels enter the separatrix zone.

If $X<0$ and $Y>0$ so that subspace $\mathcal{C}$ is shrinking along with $\mathcal{A}$, or if $X>Y>0$ so that $\mathcal{B}$ is shrinking, then only one subspace is growing, and all system state amplitude which leaves the other subspaces must emerge from the separatrix zone into the single growing subspace, because it cannot enter a shrinking one. 

If $Y>X>0$, however, then $\mathcal{A}$ is shrinking while both $\mathcal{B}$ and $\mathcal{C}$ are growing. Both growing regions are eligible to receive immigrant amplitude---and the question is how the system's state distributes itself between them. It will suffice to focus on this case $Y>X>0$, with $\mathcal{A}$ being divided among $\mathcal{B}$ and $\mathcal{C}$: the other two cases with non-trivial distribution decisions are exactly analogous.

The quantum analog of the KNH theorem, when only $\mathcal{A}$ is shrinking, would be that the probability for the system to emerge from the separatrix zone in the $\mathcal{C}$ subspace is 
\begin{equation}\label{pac1}
P_{\mathcal{A}\to\mathcal{C}}=\frac{dD_\mathcal{A}/d\lambda}{dD_\mathcal{C}/d\lambda}=\frac{X}{Y}.
\end{equation}
How well does this prediction apply to actual quantum evolution through the unitary lattice of $\hat{U}_{mn}$? As we will see below, the quantum KNH probability (\ref{pac1}) is \emph{not} correct in general for any single initial eigenstate $|E^\mathcal{A}_m\rangle \in A$. It does hold exactly as an \emph{average} probability, however, when the average is correctly (and realistically) defined.
 
\section{The quantum KNH theorems}
\begin{figure} 
\centering
\includegraphics[width =0.48\textwidth]{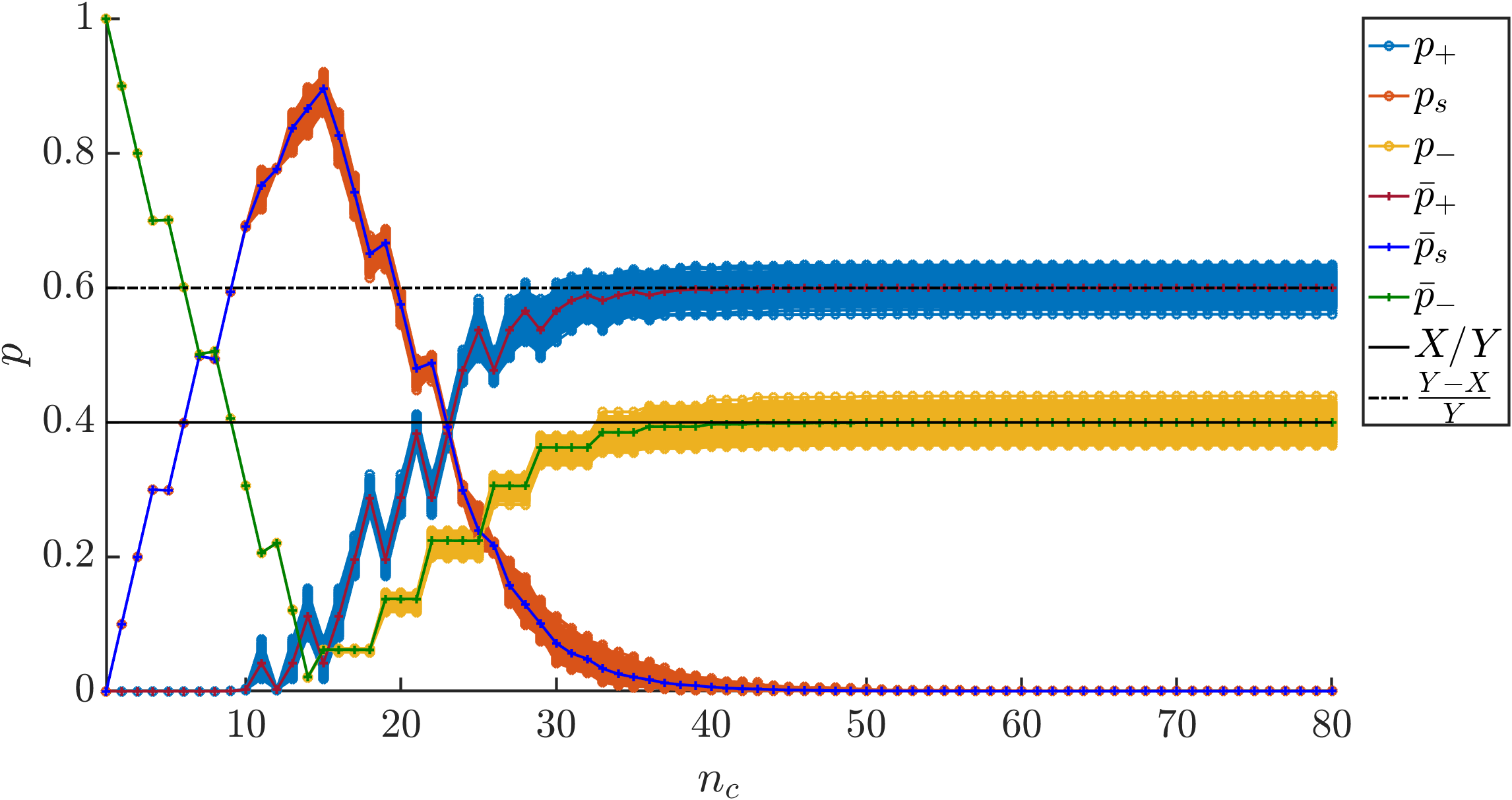}

\caption{An initial microcanonical ensemble of $M=10$ consecutive $E^\mathcal{A}_m$ eigenstates is evolved through the $\hat{U}_{mn}$ lattice; shown are the probabilities to end up above ($p_+$) or below the separatrix ($p_-$), versus time, where time is measured as the number of avoided crossings through which the system has evolved in the $\hat{U}_{mn}$ lattice. The $\hat{U}_{mn}$ are given by (\ref{Umn}) with $P_{mn}$ given by (\ref{Pmn}), with $X=0.5$, $Y=1.25$ and $Z=1$. After the ensemble has fully exited the separatrix zone into the energy ranges above and below it, the probabilities correspond to $p_+= P_{\mathcal{A}\to\mathcal{B}}$ and $p_- = P_{\mathcal{A}\to\mathcal{C}}$. The separatrix zone is interpreted as the set of crossings for which $0.001 < P_{mn} < 0.999$, and has width $D\sim 10$. Every phase $a,b,c$ in each $\hat{U}_{mn}$ is chosen randomly, independently for each crossing $(m,n)$. Initially the whole ensemble is below the separatrix; as evolution continues the ensemble enters the separatrix zone and eventually disperses above and below it. The red, yellow, and blue curves appear thick because they are the superposed curves for 100 different random realizations of all the $a,b,c$ phases in the $\hat{U}_{mn}$ lattice. The sharp $\bar{p}_{\pm,s}$ curves are the averages over the 1000 realizations; they are indistinguishable from the results when the unitary $\hat{U}_{mn}$ are replaced with classical random choices with probabilities $P_{mn}$ and $1-P_{mn}$. The averaged $\bar{p}_\pm$ for $n_c\gtrsim 50$ obey the strong quantum KNH theorem and agree exactly with its $X/Y$ and $1-X/Y$ predictions; the individual $p_\pm$ runs show fluctuations around $\bar{p}_\pm$ that are well within the $\pm D/M$ bounds of the weak quantum KNH theorem.}
\label{finalprob}\end{figure}

Fig.~\ref{finalprob} shows an example in which an initial ensemble of ten successive $E^\mathcal{A}_m<E_S$ states evolves into and through a quantum separatrix zone. The $\hat{W}$ in each $\hat{U}_{mn}$ according to (\ref{Umn}) has $P_{mn}$ given by (\ref{Pmn}) with $Z=1$, $X=0.5$ and $Y=1.25$; the three phases in all the $\hat{U}_{mn}$ are independently random. These phases should in fact all be fixed deterministically by the particular system Hamiltonian and $\lambda(t)$, but in the adiabatic limit large phases accumulate over the long times between crossings, and their values modulo $2\pi$ depend so sensitively on the precise form of $V(x,\lambda)$ and $\lambda(t)$ that they can easily be anything, and so the independent random phases used to compute Fig.~\ref{finalprob} represent a generic slowly time-dependent double well system. The curves in the Figure at $n_c = 80$ show the final probabilities $p_- = P_{\mathcal{A}\to\mathcal{C}}$ and $p_+ = P_{\mathcal{A}\to\mathcal{B}}$ to be in some $|E^\mathcal{C}_n\rangle$ state with $E^\mathcal{C}_n$ below the separatrix zone, or in any eigenstate above the separatrix zone, respectively. The results show small fluctuations, for each realization of all the random phases, around the KNH predictions of $p_- = X/Y = 0.4$ and $p_+ = 1-p_- = 0.6$. The averages over all the phase realizations match $p_-=0.4$ and $p_+ =0.6$ precisely. This Figure's example shows what the KNH theorem can mean, concretely, for a quantum system. We will now explain this example by proving the general quantum KNH theorem, first in a weak version, and then in a strong one.
  
\subsection{The weak quantum KNH theorem}
Here we show, from unitarity as the quantum analog of Liouville's theorem, that an initial microcanonical ensemble of $M$ $|E^\mathcal{A}_m\rangle$ eigenstates, initially within an adiabatically shrinking dynamical subspace, evolves through the separatrix zone into a final mixed state with probabilities to be in the two growing dynamical subspaces that are given by the KNH predictions, to within discrepancies of order $D/M$, where $D$ is the width in levels of the separatrix zone itself. This weak result is already sufficient to establish correspondence with the classical KNH theorem, since in the classical limit $\hbar\to 0$, $D$ remains finite while the number of eigenstates $M$ within a microcanonical ensemble of any classical energy width $\Delta E$ becomes infinite. The weak quantum KNH theorem will then also be used, in combination with one other physical consideration, to prove our stronger result.

The separatrix zone consists by definition of some fixed, finite number $D$ of instantaneous energy levels. The exact value of $D$ depends on how small a diabatic or adiabatic Landau-Zener amplitude we are prepared to neglect, but since these amplitudes decrease as double exponentials with energy away from the quantum separatrix energy, some finite $D$ can always be found to satisfy any desired degree of precision.

\begin{figure}
	\centering
\includegraphics[width = 0.42\textwidth]{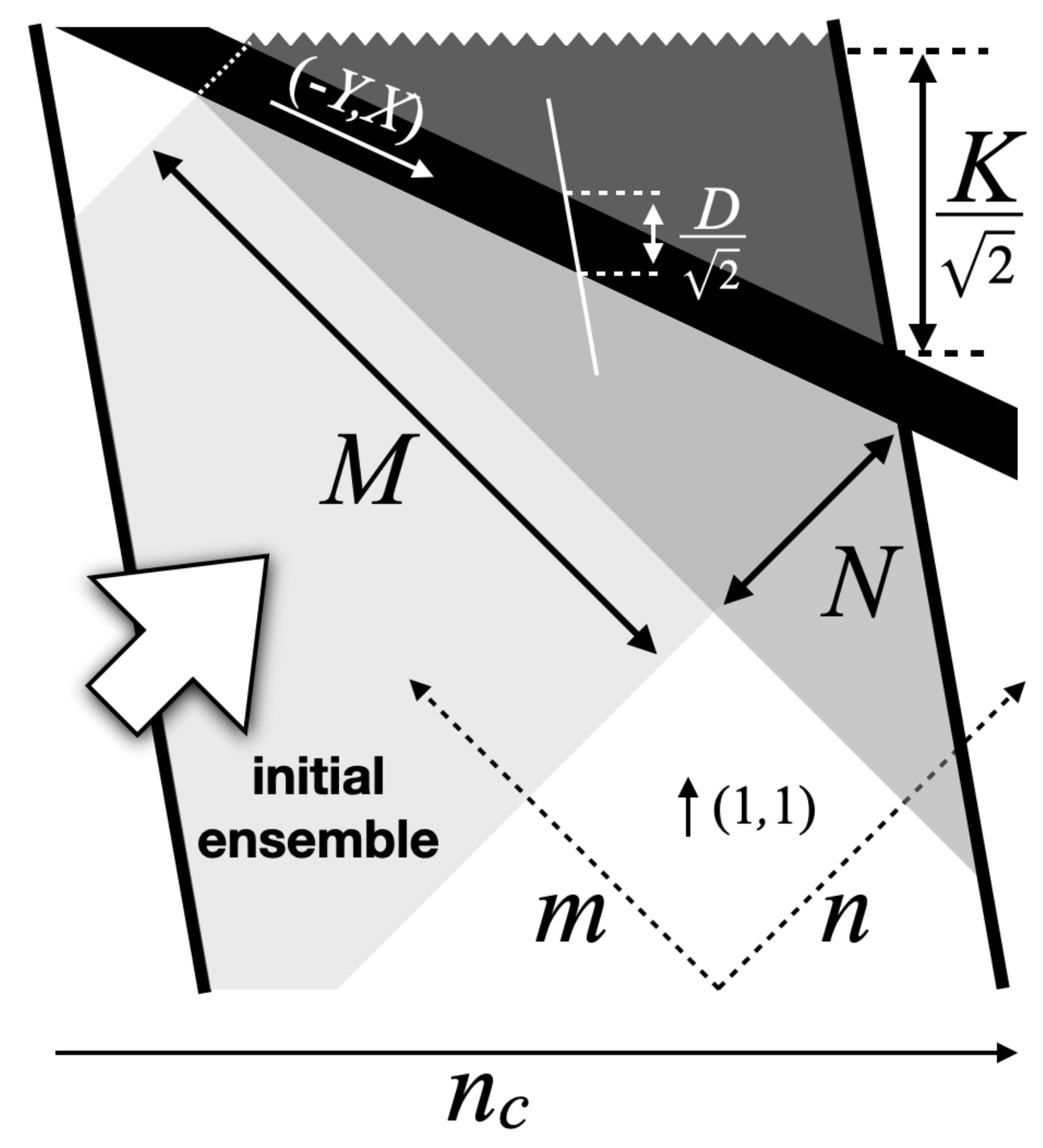}
	\caption{Evolution of an initial ensemble of $M$ successive $E^\mathcal{A}_m$ states in the $(E,\lambda)$ plane, for an illustrative case with $0<X<Y$ (region $\mathcal{A}$ shrinking, $\mathcal{B}$ and $\mathcal{C}$ both growing). The $(E,\lambda)$ plane is represented with the skewed continuous $(m,n)$ coordinates of Fig.~\ref{LatticeFig}b). The parallel thick lines on the left and right sides of the figure, tilted slightly to the left of vertical, are lines of constant $\lambda$, representing the initial time (left line) and the point $\lambda=\lambda^*$ (right line) when the initial ensemble has fully entered the separatrix zone. The wide black band in the middle of the figure is the separatrix zone. $K$ and $N$ are numbers of adiabatic eigenstates in $\mathcal{B}$ and $\mathcal{C}$ subspaces, respectively. At the later time, the density operator of the initial microcanonical ensemble of $M$ states has become a normalized projection operator of rank $M$ into a subspace of dimension $K+N+D$.}
	\label{fig:TheoremFig}
\end{figure}

The evolution within the separatrix zone is in general complicated---a quantum weighted random walk with many interfering paths and many phases---but outside the separatrix zone the evolution is by definition simple; see Fig.~\ref{fig:TheoremFig}. An initial
microcanonical ensemble of $M$ adjacent instantaneous Bohr-Sommerfeld levels $|E^\mathcal{A}_m\rangle$ that are all rising with $\lambda$ towards the separatrix must therefore move over time entirely into the separatrix zone, and at the moment (call it $\lambda=\lambda^*$) when it has fully done this, it will have evolved entirely into a Hilbert subspace of dimension $K+N+D$, where the $K$ states are in the adiabatic subspace $\mathcal{B}$ above the separatrix zone, the $N$ states are in the subspace $\mathcal{C}$ of diabatically crossing $E^\mathcal{C}_n$ levels below the separatrix zone, and the $D$ states are inside the separatrix zone. No evolution outside this subspace is possible up to the time when $\lambda(t)=\lambda^*$, because of the strong constraints of essentially perfect adiabaticity/diabaticity outside the separatrix zone. We will refer to the $K$ possibly populated levels in $\mathcal{B}$ just above the separatrix zone at $\lambda^*$ as the $K$-dimensional subspace $\mathcal{B}^*$, to the $N$ possibly populated levels in $\mathcal{C}$ just below the separatrix zone at $\lambda^*$ as $\mathcal{C}^*$, so that the total subspace that can be populated at $\lambda^*$ is $\mathcal{B}^*\cup \mathcal{C}^*\cup\mathcal{S}^*$, where $\mathcal{S}^*$ is the separatrix zone at $\lambda(t)=\lambda^*$.

Fig.~\ref{fig:TheoremFig} illustrates the resulting geometrical relationships between the separatrix parameters $X,Y$ and the subspace dimensions $D,K,N$. See \ref{app:lattice_geometry} for a detailed explanation; the result is
\begin{align}
\label{KN_geom}
    N &= M\frac{X}{Y} + \delta N\nonumber\\
    K &= (M-kD)\Big(1-\frac{X}{Y}\Big) + \delta K\;,
\end{align}
where $|\delta N|<1/2$ and $|\delta K|<1/2$ come from exactly how the discrete lattice of integer $(m,n)$ lines up with the real-number slope $X/Y$, to make $K$ and $N$ be integers. The constant $k$ depends on $X$ and $Y$, and on the ratio $\partial_E\tilde{S}_\mathcal{A}/\partial_E\tilde{S}_\mathcal{C}$ that determines the slope of a line of constant $\lambda$ in the $(m,n)$ plane; the $-kD$ term is present in $K$ because there is some delay between when the first $|E^\mathcal{A}_m\rangle$ state of the initial ensemble enters the separatrix zone, and when amplitude from it begins to emerge into the $\mathcal{B}$ subspace, having crossed the separatrix zone at the maximum rate of one level per crossing. The exact value of $k$ is not important for the weak quantum KNH theorem, which concerns the limit $D/M\to 0$.

The density operator $\hat{\rho}_I$ for the initial microcanonical ensemble is an identity operator of rank $M$, divided by $M$, and the evolution is unitary. The state at $\lambda(t)=\lambda^*$ is therefore
\begin{equation}
    \hat{\rho}_{\lambda^*} = \frac{1}{M}\sum_{m=1}^M |\Psi_m\rangle\langle\Psi_m|
\end{equation}
for some set of states $|\Psi_m\rangle$ in the final Hilbert subspace of dimension $K+N+D = M +\mathcal{O}(D)$. The probability to be in any of the three subspaces $\sigma=\mathcal{B}^*,\mathcal{C}^*,\mathcal{S}^*$ that are populated at $\lambda^*$ is
\begin{equation}
    P^*_{\sigma} = \mathrm{Tr}\big(\hat{\rho}_{\lambda^*}\hat{\Pi}_{\sigma}\big)
\end{equation}
where $\hat{\Pi}_{\sigma}$ are the projection operators onto the subspaces $\sigma=\mathcal{B}^*,\mathcal{C}^*,\mathcal{S}^*$, which satisfy $\hat{\Pi}_{\mathcal{B}^*}+\hat{\Pi}_{\mathcal{C}^*}+\hat{\Pi}_{\mathcal{S}^*}=\hat{I}^*$ when $\hat{I}^*$ is the projector onto the total space $\mathcal{B}^*\cup \mathcal{C}^* \cup \mathcal{S}^*$. Using the triangle-related inequality
\begin{equation}
\mathrm{Tr}\big(\hat{\rho}_{\lambda^*}\hat{\Pi}_\sigma\big) \leq \frac{D^*_\sigma}{M}
\end{equation}
where $D^*_\sigma = K,N,D$ is the rank of $\hat{\Pi}_\sigma$, and the identity $\mathrm{Tr}\big(\hat{\rho}_{\lambda^*}\hat{I}^*\big)\equiv 1$, we can establish the inequalities
\begin{align}
\label{lamstarQKNH}
    P^*_\mathcal{B} &= \mathrm{Tr}\Big[\hat{\rho}_{\lambda^*}\big(\hat{I}^*-\hat{\Pi}_{\mathcal{C}^*}-\hat{\Pi}_{\mathcal{S}^*}\big)\Big]\nonumber\\
    &\geq 1-\frac{N+D}{M}=1-\frac{X}{Y}-\mathcal{O}(D/M)\\
    P^*_\mathcal{C} &= \mathrm{Tr}\Big[\hat{\rho}_{\lambda^*}\big(\hat{I}^*-\hat{\Pi}_{\mathcal{B}^*}-\hat{\Pi}_{\mathcal{S}^*}\big)\Big]\nonumber\\& \geq 1-\frac{K+D}{M} = \frac{X}{Y} - \mathcal{O}(D/M)\;.
\end{align}

Further evolution to times later than $\lambda(t)=\lambda^*$ can never lower either $P_\mathcal{B}$ below $P^*_\mathcal{B}$ or $P_\mathcal{C}$ below $P^*_\mathcal{C}$, because the respectively diabatic and adiabatic crossings in $\mathcal{B}$ and $\mathcal{C}$ only bring adiabatic eigenstates further into the $\mathcal{B}$ and $\mathcal{C}$ subspaces. $P_\mathcal{B}$ and $P_\mathcal{C}$ can only increase above their values at $\lambda^*$, as probability that is still in $\mathcal{S}$ at $\lambda^*$ migrates out of $\mathcal{S}$ into $\mathcal{B}$ and $\mathcal{C}$. In fact, all of the system's amplitude to be in the separatrix zone subspace $\mathcal{S}$ must eventually leave $\mathcal{S}$, bringing $P_\mathcal{S}\to 0$, because within the separatrix zone we have both $P_{mn}<1$ and $1-P_{mn}<1$. In the final state after all amplitude has emerged from $\mathcal{S}$ into $\mathcal{B}$ and $\mathcal{C}$, therefore, we must have $P_\mathcal{B} + P_\mathcal{C} = 1$. This yields the \emph{weak quantum KNH theorem} for the probabilities $P_{\mathcal{A}\to \mathcal{B,C}}$ when $\mathcal{B}$ and $\mathcal{C}$ are both growing:
\begin{align}
    P_{\mathcal{A}\to\mathcal{C}} &\geq \frac{X}{Y}-\mathcal{O}(D/M) = 1-P_{\mathcal{A}\to\mathcal{B}}\nonumber\\&\leq \frac{X}{Y}+\mathcal{O}(D/M)\nonumber\\\label{WQKNH}
    \Longrightarrow\; P_{\mathcal{A}\to\mathcal{C}} &= \frac{X}{Y} \pm \mathcal{O}(D/M)\;. 
\end{align}
The analogous results can be shown similarly for the other cases in which two of the three energy subspaces are growing with $\lambda$.

The weak quantum KNH theorem is weak in the sense that (\ref{WQKNH}) allows a margin of error $\mathcal{O}(D/M)$. As we have explained above, this suffices to establish quantum-classical correspondence of the KNH theorem, because in the limit $\hbar\to0$ we have $M\to\infty$ for any fixed energy width while $D$ remains finite.  As in the case of Fig.~\ref{finalprob}, however, Eqn.~(\ref{WQKNH}) can easily be much more generous than necessary. Our derivation of (\ref{WQKNH}) has relied only on unitarity and on the simple forms of evolution outside the separatrix zone; we have not even attempted to analyse the complicated quantum random walk which occurs inside the separatrix zone $\mathcal{S}$. Consequently our bounds on $P_{\mathcal{A}\to\mathcal{C}}-\frac{X}{Y}$ have had to allow the most pessimistic scenario, in which all of the probability which is still in $\mathcal{S}$ at $\lambda^*$ might finally move into either $\mathcal{B}$ or $\mathcal{C}$. Complicated as it may be, the unitary evolution inside $\mathcal{S}$ is clearly not actually going to be so arbitrary; the separatrix zone is defined, after all, by the fact that it has no simple bias in how it distributes amplitude. 

Since the random walk inside $\mathcal{S}$ will take $\mathcal{O}(D)$ steps before $P_\mathcal{S}$ falls to an arbitrarily low level, we might expect the typical difference $P_{\mathcal{A}\to\mathcal{C}}-\frac{X}{Y}$ to be of order $\sqrt{D}/M$ for large $D$, well below the strict upper limit of $D/M$. This scaling does seem to be consistent with numerical experiments for different realizations of the $\hat{U}_{mn}$ lattice with independent random phases in the $\hat{U}_{mn}$. Computing a rigorous prefactor for the $\sqrt{D}/M$ correction to $P_{\mathcal{A}\to\mathcal{C}}$ for general $\hat{U}_{mn}$ would seem to be difficult, however; many arbitrarily different unitary phases must be admitted, and many complicated paths through the lattice of crossings must all be allowed to interfere quantum mechanically. Even without being able to solve that problem, however, we can show that the precise KNH result $P_{\mathcal{A}\to\mathcal{C}} = X/Y$ for the average over many random phase realizations in Fig.~\ref{finalprob} was by no means an accident.

\subsection{The strong quantum KNH theorem}
In the adiabatic limit for which all KNH results hold, the phases which all quantum amplitudes acquire during the long periods between avoided crossings are large. Modulo $2\pi$, therefore, they are effectively random, even though they are strictly determined by $V(x,\lambda(t))$, in the sense that arbitrarily small changes to $V(x,\lambda)$ or to $\lambda(t)$ could make these phases arbitrarily different. What this implies for experiments is that the three phases $a,b,c$ in each $\hat{U}_{mn}$ are not actually reproducible: they will inevitably be independently random in every run of any series of experiments. An experimental measurement of $P_{\mathcal{A}\to\mathcal{C}}$ will therefore not actually probe the coherent random walk of $\hat{U}_{mn}$, but only the classically probabilistic weighted random walk, with probabilities $P_{mn}$ and $1-P_{mn}$ at each crossing, that results from averaging over all the $a,b,c$ phases at each $(m,n)$.
\begin{figure*}
\centering
    \includegraphics[scale=0.5]{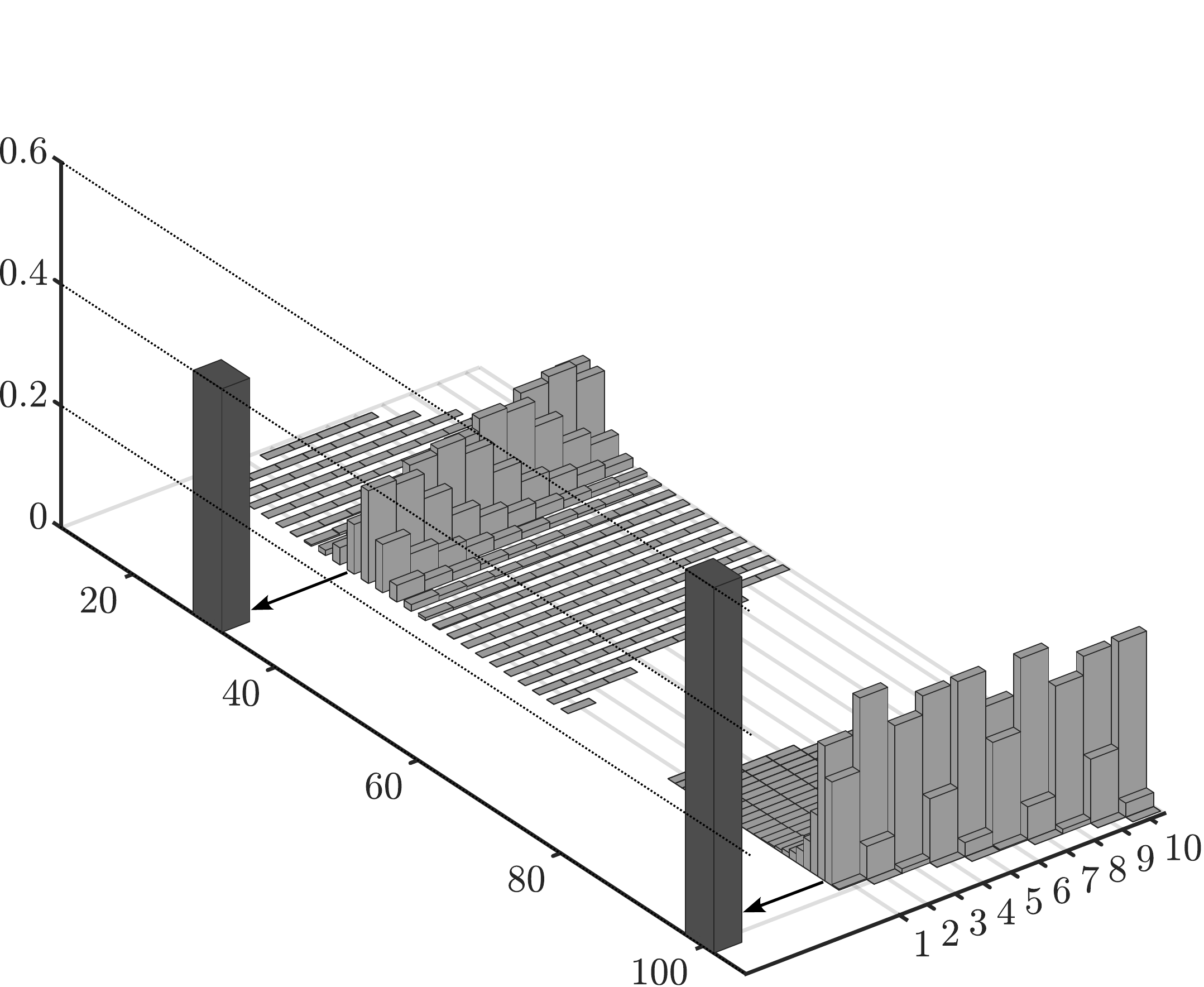}
 \caption{Distribution of final adiabatic states for $M=10$ adjacent rising initial states after $n_c =80$ incoherent Landau-Zener crossings, with probabilities given by $P_{mn}$ from Eqn.~(\ref{Pmn}) with $Z=1$, $X=0.5$, $Y=1.25$. The evolution is computed with probabilities, instead of unitary evolution with independent random phases, since this yields the probability distribution which would be inferred from any realistic set of experiments, in which the large adiabatic phases between each avoided crossing are not reproducible modulo $2\pi$ from run to run. The final state with the number 1 is the lowest final state that can be reached ($n_c-1$ levels below the lowest initial state of the ensemble) by following the diabatic path after the first avoided crossing. The initial states with labels 1 to 10 are the state that is directly below the separatrix zone at the first avoided crossing and the 9 adjacent states below it. The heights of the \textbf{light gray columns} in each horizontal row represent the probability to emerge from the separatrix zone into various final states, from an initial state with energy eigenvalue below the quantum separatrix $E^\mathcal{A}_m < E_s$, which over time rises adiabatically into the separatrix zone. Successive horizontal rows refer to different initial $E^\mathcal{A}_m$, for ten consecutive $m$. The final states are $E^\mathcal{C}_n$ below the separatrix, while above the separatrix they alternate, with increasing energy, between $E^\mathcal{A}_m$ and $E^\mathcal{C}_n$, since above the separatrix the system switches adiabatically between $E^\mathcal{A}_m$ and $E^\mathcal{C}_n$ at each avoided crossing. Final states within the separatrix zone are not shown, since their probabilities are all negligible after eighty crossings. The \textbf{dark gray columns} indicate the total probability to emerge anywhere below the separatrix (left dark column) or anywhere above it (right dark column), averaged over all 10 initial states. While the light gray columns show a complicated pattern, the heights of the dark gray columns are nevertheless in exactly the ratio $X:Y$ that is deduced from unitarity by the strong quantum KNH theorem.}
\label{numevol}\end{figure*}

With this additional physical insight that only the $P_{mn}$ are observable in the adiabatic limit, we can prove a stronger quantum KNH theorem from the weak theorem, by considering the hypothetical case in which (\ref{Pmn}) holds not only for $m,n\lesssim \mathcal{O}(\hbar^{-1})$, but for arbitrarily large $m$ and $n$.

Suppose that $X/Y$ is a rational number $q/p$ for some minimal integers $q,p$; even irrational $X/Y$ can be approximated arbitrarily closely by such rationals, and so as far as any experiments are concerned the assumption that $X/Y = q/p$ can be made without loss of generality. The pattern of probabilities (\ref{Pmn}) therefore repeats itself exactly for $P_{m+p,n-q}$. Consider, then, the purely hypothetical case of a lattice of Landau-Zener probabilities that extends to infinite $m,n$ with perfect regularity, and has this same periodicity $P_{m+p,n-q} = P_{m,n}$.

In this purely hypothetical case consider an initial ensemble of $M_J=q\times J$ contiguous $E^\mathcal{A}_m$ levels, where $J$ is an arbitrarily large integer. Because of the lattice periodicity of $P_{mn}$, the total $P_{\mathcal{A}\to\mathcal{C}}$ of this large ensemble for any $J>1$ must be exactly the same as we would find for the narrower ensemble with $J=1$ and in the non-hypothetical case where (\ref{Pmn}) is only valid over a range $m,n \lesssim q + \mathcal{O}(D)$. By the weak quantum KNH theorem (\ref{WQKNH}), however, we have for the hypothetically extended $m,n$-range of (\ref{Pmn}) and the $M_J$-sized ensemble
\begin{equation}\label{SQKNH}
   P_{\mathcal{A}\to\mathcal{C}} =  \lim_{J\to\infty}P_{\mathcal{A}\to\mathcal{C}} = \frac{X}{Y}\;.
\end{equation}

Consequently even for $J=1$ we have $P_{\mathcal{A}\to\mathcal{C}} = X/Y$, with \emph{no} $\mathcal{O}(D/M)$ margin of error. This explains the perfect agreement of the late-time probabilities in Figs.~\ref{numevol} and \ref{finalprob}, which is a case with $(q,p)=(2,5)$ and $J=2$. 

We name (\ref{SQKNH}) the \emph{strong quantum KNH theorem} for cases $0<X<Y$ (only $\mathcal{A}$ is shrinking); other cases were resolved in III.C, above, as either $P_{\mathcal{A}\to\mathcal{C}}=0$ or $P_{\mathcal{A}\to\mathcal{C}}=1$. As Fig.~\ref{numevol} shows, the result (\ref{SQKNH}) does \emph{not} apply to the final probabilities that evolve from any single initial energy eigenstate; even after eliminating quantum interference in the random walk through the separatrix zone by averaging over effectively random phases, the doubly exponential $P_{mn}$ of (\ref{Pmn}) provide a formidably complicated set of decision weights, and many different paths through the lattice still contribute to the final probabilities. The strong quantum KNH theorem is a kind of sum rule, however, which strictly governs the average probability when the initial energy cannot be exactly controlled.

For any initial ensemble which is not microcanonical with width in levels equal to an integer multiple $J$ of the lattice periodicity of $P_{mn}$, the probabilities to end up in different dynamical regions of Hilbert space may not satisfy the strong quantum KNH theorem, but only the weak one, or an intermediate version in which the correction is determined by the mismatch between $M$ and $Jq$. Another way of expressing the strong quantum KNH theorem, however, is to say that for randomly selected initial energy eigenstates the average probability to emerge from the separatrix zone in the different possible subspaces is indeed given exactly by the quantum KNH result (\ref{SQKNH}). 

\section{Summary and discussion}
In conclusion we summarize our main results. For a slowly time-dependent Hamiltonian similar to that of a double-well potential, the classical concept of a separatrix does extend smoothly to quantum mechanics, in the form of a narrow range of energies within which Landau-Zener transitions at each avoided level crossing are intermediate between diabatic and adiabatic. The quantum separatrix generally has lower energy than the barrier height, by an amount that depends on the rate at which the potential is changing. In the semi-classical limit where the time-dependent energy levels form a locally regular lattice, the Landau-Zener probabilities in the separatrix zone take a universal form defined by three real numbers $X$, $Y$ and $Z$, where $Z$ fine-adjusts the overall position of the discrete lattice relative to the quantum separatrix energy, and $X$ and $Y$ determine the width of the separatrix zone as well as its slope in the $(E,\lambda)$ (or $(m,n)$) plane. Depending on this slope, the dynamically distinct subspaces into which the separatrix divides the quantum energy spectrum may all be growing or shrinking in time. 

First of all we found that no amplitude can migrate into a shrinking subspace. Then from unitarity and geometry we could derive the weak quantum KNH theorem, relating probabilities to emerge from the separatrix zone in different growing subspaces to their rates of growth in dimensionality as given by $X$ and $Y$, within error bounds of order $D/M$. Finally we could use the periodicity of the $P_{mn}$ lattice and the weak theorem to prove the strong quantum KNH theorem, which applies to probabilities averaged over adiabatically irreproducible phases and has error bounds of zero for initial ensembles which match the period of the $P_{mn}$ lattice.

Together these results extend into quantum mechanics the KNH connection between probabilities of post-adiabatic change and growth rates of phase space areas. This connection may offer a microscopic basis for the Second Law of Thermodynamics which does not depend on assumptions about equilibration, inasmuch as the areas which must grow according to KNH theorems do not have to be explored ergodically by the system.

\subsection{The quantum KNH theorem beyond the semi-classical limit}
In cases where $\hbar$ is not so small in comparison with the classical action scales in the problem, so that the range of validity $|m|,|n|<\mathcal{O}(\hbar^{-1})$ of our key equation (\ref{Pmn}) for the Landau-Zener diabatic probability $P_{mn}$ is not very broad, the range of $m,n$ over which our $P_{mn}$ is periodic may exceed the range over which it is valid. In this case the strong quantum KNH theorem will not hold. And if the $\mathcal{O}(\hbar^{-1})$ range of $m$ and $n$ within which (\ref{Pmn}) holds is not even wide enough to cover an initial ensemble width $M\gg D$, then even the weak quantum KNH theorem may set only rather loose unitarity bounds on the probabilities with which the system emerges on different sides of the separatrix zone. 

For the validity as a concept of the quantum separatrix zone, however, the semi-classical limit is only a sufficient condition, not a necessary one. If a quantum system evolving under slowly time-dependent conditions features avoided level crossings with avoidance width changing sufficiently quickly with energy---for whatever reason---then the Landau-Zener transition probabilities at these crossings will have a correspondingly abrupt crossover from diabatic to adiabatic.  An equation similar to our (\ref{Pmn}) may then be valid, implying that unitarity will constrain transition probabilities through this narrow separatrix zone in the energy spectrum, with $X/Y$-like rules that are directly analogous to the KNH rule for classical separatrix crossing. For quantum systems outside the WKBJ limit we must simply replace the classical phase space areas with the numbers of quantum energy levels $D_\mathcal{A,B,C}$ that are contained within each subspace that is defined by the quantum separatrix.

Since the two measures of phase space area and energy subspace dimension coincide in the WKBJ limit, the WKBJ semiclassical theory is once again providing its usual bridge between quantum and classical dynamics. Even though classical adiabaticity breakdown at the classical separatrix energy (the barrier height) does not extend to quantum mechanics, our quantum generalization of the separatrix concept preserves this important qualitative feature of classical mechanics. The quasi-classical KNH behavior then emerges, ironically, via a quantum random walk through the lattice of quantized energy levels with crossings that are narrowly avoided because of quantum tunnelling.

\subsection{The quantum KNH theorem beyond the double well}
Both classically and quantum mechanically, more complicated separatrices are possible that will divide phase space or the energy spectrum into more than three regions. A multi-well system is an obvious example. The principles of the classical and quantum KNH theorems apply in these cases but their detailed implications may not be trivial and will require further study.

Even with only three distinct dynamical regions, the two regions that overlap in energy do not necessarily have to be lower in energy than the third region. It can be the other way around, for example in the case of a pendulum, where there is only one lower-energy region of back-and-forth oscillation, with two degenerate high-energy regions of full rotation in clockwise or counter-clockwise directions. As we will report in more detail in future, the quantum KNH theorem that we have developed here for double-well-like systems applies to such pendulum-like systems as well, with an inversion of energy. The quantum separatrix lies in general \emph{above} the classical separatrix; quantum transitions between the classically separate forms of higher-energy dynamics are provided here, not by tunnelling through a potential barrier, but by non-classical above-barrier reflection. A precisely similar lattice of Landau-Zener transitions results, with precisely analogous quantum KNH behavior emerging in a precisely similar way. In the semi-classical limit the double-well tunnelling exponents $T_b$ even have counterparts, for above-barrier reflection in the pendulum system, that are also action integrals defined with analytically continued phase space variables.
\section{Acknowledgments}
The authors acknowledge support from State Research Center OPTIMAS and the Deutsche Forschungsgemeinschaft (DFG) through SFB/TR185 (OSCAR), Project No. 277625399. 

\section*{Appendix}
\appendix
\section{The semi-classical double well}
\subsection{Modified Bohr-Sommerfeld quantization}\label{app:bs}
As is well known, the WKBJ semi-classical approximation breaks down at classical turning points $V(x_n)=E$, where the WKBJ eikonal approximation must be supplemented by a connection formula \cite{Miller_Good}. Connection formulas are obtained by exploiting the fact that, when the relevant classical action scales are large compared to $\hbar$, WKBJ only fails within a small neighborhood of the turning point. Within this narrow range of $x-x_n$, and indeed to some distance outside it, the exact potential can be approximated well with a simpler function, for which the two independent solutions to the time-independent Schr\"odinger equation can be found exactly. One then uses the method of matched asymptotics \cite{lagerstrom1988matched} to infer, from these two locally exact solutions, the modified continuity condition which relates the amplitudes of the two WKBJ solutions that must appear, on either side of the breakdown region, in any global solution.

The best-known form of connection formula applies when oscillating WKBJ solutions within a single potential well must be connected to exponentially decaying WKBJ solutions under the potential walls on either side of the well. One approximates the potential near the turning point with a linear gradient, so that the locally exact solutions within the linearized breakdown region are Airy functions. When the resulting conditions on the WKBJ linear combinations are applied on both sides of the well, the two sets of conditions can only be satisfied simultaneously for certain discrete values of the energy $E$. This condition is precisely the Bohr-Sommerfeld quantization condition: the quantum energies $E_k$ are those of classical orbits which enclose phase space area $\pi(k+1/2)\hbar$ for integer $k$.

\begin{figure}
    \centering
    \includegraphics[width = 0.48\textwidth]{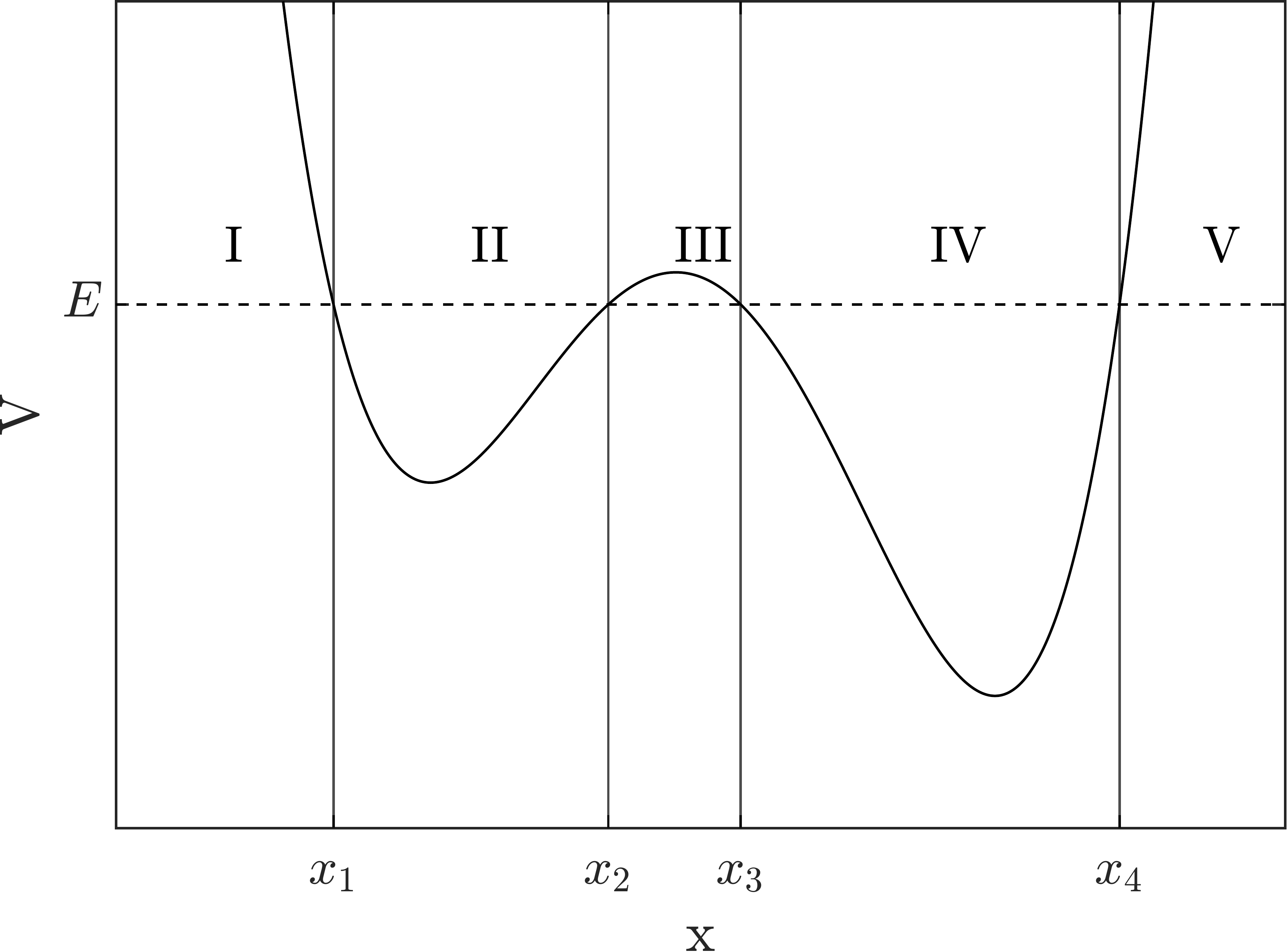}
    \caption{A generic double well potential with energy below the top of the inner barrier but close to it. There are four turning points $x_1-x_4$. The coefficients of the WKB solutions in the regions $I,II$ and $IV,V$ are related by connection formulas based on Airy-functions while the coefficients of the wave function in the region $IV$ can be connected to the coefficients of the solution in $II$ across the barrier with a connection formula based on parabolic cylinder functions \cite{CHILD1974280}. Successively applying these connection formulas and demanding that the wave function vanishes at infinity results in the modified Bohr-Sommerfeld rule (\ref{GenBS}) in our main text.}
    \label{fig:app:dw}
\end{figure}

In the more complicated case of a double well, however, there are always two outer turning points, but for $E<V_b$ there are also two inner turning points, on either side of the central potential barrier ( see Fig.\ref{fig:app:dw}). To account for tunneling, we must allow both growing and decaying solutions under the central barrier; sufficiently close to the top of the barrier, moreover, we cannot linearize $V(x)$ but must approximate it instead as a downward parabola, so that the local solutions in the WKBJ breakdown region are no longer Airy functions, but parabolic cylinder functions. 

In the following, we will briefly review how connection formulas based on Airy- and parabolic cylinder functions have to be combined in order to get the modified Bohr-Sommerfeld rule (\ref{GenBS}) as has been shown in \cite{CHILD1974280}.
The WKB solutions in the respective regions are given by
\begin{widetext}
\begin{equation}
    \psi(x) = \frac{1}{\sqrt{p(x)}}\begin{cases}
    A_{I}e^{\frac{1}{\hbar}\int_{x_1}^{x}\mathrm{d}x'\left|p(x')\right|}+B_{I}e^{-\frac{1}{\hbar}\int_{x_1}^{x}\mathrm{d}x'\left|p(x')\right|} & x < x_1\\
     A_{II}e^{\frac{i}{\hbar}\int_{x_1}^{x}\mathrm{d}x' p(x')}+B_{II}e^{-\frac{i}{\hbar}\int_{x_1}^{x}\mathrm{d}x' p(x')} & x_1 < x< x_2\\
        A_{IV}e^{\frac{i}{\hbar}\int_{x_3}^{x}\mathrm{d}x' p(x')}+B_{IV}e^{-\frac{i}{\hbar}\int_{x_3}^{x}\mathrm{d}x' p(x')} & x_3 < x < x_4\\
        A_{V}e^{\frac{1}{\hbar}\int_{x_4}^{x}\mathrm{d}x'\left|p(x')\right|}+B_{V}e^{-\frac{1}{\hbar}\int_{x_4}^{x}\mathrm{d}x\left|p(x')\right|} & x > x_4
    \end{cases}
\end{equation}
\end{widetext}
with $p(x) = \sqrt{2\mu\left(V(x)-E\right)}$. The coefficients in the respective regions are related by connection formulas as will be shown in the following. 
The coefficients $(A_I,B_I)$ are related to the coefficients $(A_{II},B_{II})$ by a connection formula for a downwards sloping turning point based on Airy functions \cite{CHILD1974280}:
\begin{equation}
    \begin{pmatrix}
    A_{II}\\B_{II}
    \end{pmatrix} = \begin{pmatrix}
    e^{-\frac{i\pi}{4}}&\frac{1}{2}e^{\frac{i\pi}{4}}\\e^\frac{i\pi}{4}&\ \ \frac{1}{2}e^{-\frac{i\pi}{4}}
    \end{pmatrix}\begin{pmatrix}
    A_{I}\\B_{I}
    \end{pmatrix}
\end{equation}
and we need to demand $B_{I} = 0$ to ensure $\psi(x\rightarrow -\infty)\rightarrow 0$. Region $III$ is assumed to be small and we want to connect the coefficients of region $II$ across the barrier with the coefficients of region $IV$ by using a connection formula based on parabolic cylinder functions \cite{CHILD1974280}:
\begin{equation}
\label{cylinderCF}
        \begin{pmatrix}
    A_{IV}\\B_{IV}
    \end{pmatrix} = \begin{pmatrix}
    \sqrt{1+e^\frac{2T_b}{\hbar}}e^{-i\Phi}&-ie^\frac{T_b}{\hbar} e^{i\theta}\\ie^\frac{T_b}{\hbar} e^{-i\theta}&\sqrt{1+e^\frac{2T_b}{\hbar}}e^{i\Phi}
    \end{pmatrix}\begin{pmatrix}
    \tilde{A}_{II}\\\tilde{B}_{II}
    \end{pmatrix}
\end{equation}
with
\begin{equation}
\label{phaseshift}
    \Phi(E,\lambda) =\left[ \arg\Gamma\left[\frac{1}{2}-\frac{iT_b}{\pi\hbar}\right] + \frac{T_b}{\pi\hbar}\left(\ln\left|\frac{T_b}{\pi\hbar}\right|-1\right)\right]
\end{equation}
where $T_b$ is the tunneling integral given by (\ref{Tb}) and $\theta = 0$ for the case $E < V_b$ we are considering here. In order to derive this connection formula, one needs to derive the connection formula for a quadratic barrier based on parabolic cylinder functions and map the general double well potential near the top of the barrier $V_b$ to a quadratic barrier  by using a turning point correspondence equation (see \cite{child2014semiclassical,Miller_Good}). The coefficients $\tilde{A}_{II},\tilde{B}_{II}$ are related to the coefficients $A_{II}, B_{II}$ by a factor that changes the phase reference point of the wave function in region $II$ to the inner left inner turning point $x_2$ as required by the connection formula \ref{cylinderCF} \cite{CHILD1974280}:
\begin{equation}
    \begin{pmatrix}
    \tilde{A}_{II}\\\tilde{B}_{II}
    \end{pmatrix} = \begin{pmatrix}
    e^{\frac{i}{\hbar}\int_{x_1}^{x_2}dx' p(x')}&0\\0&e^{-\frac{i}{\hbar}\int_{x_1}^{x_2}dx'p(x')}
    \end{pmatrix}\begin{pmatrix}
    A_{II}\\B_{II}
    \end{pmatrix}
\end{equation}
The coefficients of the wave function in region $V$ are related to the coefficients in region $IV$ by a connection formula for an upward sloping turning point based on Airy functions \cite{CHILD1974280}:
\begin{equation}
    \begin{pmatrix}
    A_{V}\\B_{V}
    \end{pmatrix} = \begin{pmatrix}
    e^{\frac{i\pi}{4}}&e^{-\frac{i\pi}{4}}\\\frac{1}{2}e^\frac{-i\pi}{4}&\frac{1}{2}e^{\frac{i\pi}{4}}
    \end{pmatrix}\begin{pmatrix}
    \tilde{A}_{IV}\\\tilde{B}_{IV}
    \end{pmatrix}
\end{equation}
where the phase reference points of the wave functions in regions $IV,V$ must be matched by
\begin{equation}
    \begin{pmatrix}
    \tilde{A}_{IV}\\\tilde{B}_{IV}
    \end{pmatrix} = \begin{pmatrix}
    e^{\frac{i}{\hbar}\int_{x_3}^{x_4}dx' p(x')}&0\\0&e^{-\frac{i}{\hbar}\int_{x_3}^{x_4}dx'p(x')}
    \end{pmatrix}\begin{pmatrix}
    A_{IV}\\B_{IV}
    \end{pmatrix}
\end{equation}
in order to apply the connection formula. Applying these connection formulas successively leads to the coefficient
\begin{align}
    A_V = 2A_I\Bigg(&\sqrt{1+e^\frac{2T_b}{\hbar}}\cos\Big(\frac{S_A+S_C}{\hbar}-\Phi\Big) \nonumber\\&+ e^\frac{T_b}{\hbar}\cos\left(\frac{S_A - S_C}{\hbar}\right)\Bigg)
\end{align}
with $S_{A,C}$ given by 
\begin{align}
    S_\mathcal{A}(E,\lambda)&=\int_{x_1(E,\lambda)}^{x_2(E,\lambda)}\!dx\,\sqrt{2\mu[E-V(x,\lambda)]}\nonumber\\
    S_\mathcal{C}(E,\lambda)&=\int_{x_3(E,\lambda)}^{x_4(E,\lambda)}\!dx\,\sqrt{2\mu[E-V(x,\lambda)]}
\end{align}
Demanding $A_V = 0$ to ensure $\psi(x\rightarrow\infty)\rightarrow0$ results in the modified Bohr-Sommerfeld rule (\ref{GenBS}).
The modified actions $\tilde{S}_\mathcal{A,C}$ that appear in (\ref{GenBS}) are
\begin{equation}
    \tilde{S}_\mathcal{A,C}(E,\lambda)=S_\mathcal{A,C}(E,\lambda)-\frac{\hbar}{2}\Phi(E,\lambda)
\end{equation}
with the phase shift \cite{CHILD1974280} due to the connection with parabolic cylinder functions being (\ref{phaseshift}). 
In addition to being valid for energies that are only slightly below the top of the barrier $V_b$, because it is based on parabolic cylinder functions and a quadratic potential rather than Airy function in a linear potential, for small $e^{-T_b/\hbar}$ (\ref{GenBS}) reduces smoothly to the result one obtains for lower energies, by using separate Airy connections at both inner turning points, with decaying and growing WKBJ solutions inside the barrier. In any semi-classical regime, therefore, it is safe to use the modified Bohr-Sommerfeld condition (\ref{GenBS}) for all $E<V_b$.

\textit{Non-perturbative accuracy}. A subtle point is that (\ref{GenBS}) is valid to leading order in $e^{-T_b/\hbar}$ even though, as a WKBJ result, it is also subject to corrections of order $\hbar$ that may be much \emph{larger} than $e^{-T_b/\hbar}$. The reason is that the post-semi-classical order $\hbar$ corrections are \emph{multiplicative}, being of the form $\times[1+\mathcal{O}(\hbar)]$. This is important for the Landau-Zener theory of avoided crossings, because the minimum energy gap at each avoided crossing is really zero to all orders in $\hbar$; its actual non-zero width $\propto e^{-T_b/\hbar}\times[1+\mathcal{O}(\hbar)]$ is \emph{non-perturbative} in $\hbar$, and its leading term is really correctly given by (\ref{GenBS}) even though (\ref{GenBS}) has $\mathcal{O}(\hbar)$ corrections and it may well be that $e^{-T_b/\hbar}\ll \mathcal{O}(\hbar)$.

\textit{Quantum adiabaticity near $E=V_b$}.
When the potential takes the form 
\begin{equation}
V(x,\lambda)\to V_b(\lambda) - \frac{\kappa(\lambda)}{2}\big(x-x_0(\lambda)\big)^2+\mathcal{O}(x-x_0)^3
\end{equation}
near the top of the barrier, for some constant $\kappa$, the classical actions behave as
\begin{equation}
   \lim_{E\to V_b^-} S_\mathcal{A,C} = S_\mathcal{A,C}(V_b,\lambda) - \frac{V_b-E}{2\sqrt{\kappa/\mu}}\ln \frac{\bar{V}_\mathcal{A,C}}{V_b-E}\,
\end{equation}
where $\bar{V}_\mathcal{A,C}$ are energy scales which depend on $V(x,\lambda)$ over the left and right wells, respectively. (Recall that $\mu$ is the particle mass.) The orbital period in each well is $2\pi (\partial S_\mathcal{A,C}/\partial E)^{-1}$, which diverges logarithmically as $E\to V_b^-$, implying the breakdown of classical adiabaticity close to the separatrix. Since
\begin{equation}
    \lim_{E\to V_b^-} T_b(E) = \sqrt{\frac{\mu}{\kappa}}(V_b-E)\;,
\end{equation}
however, the shifted actions $\tilde{S}_\mathcal{A,C}$ which appear in (\ref{GenBS}) remain smooth through the separatrix, with the $\ln(V_b-E)$ terms canceling, providing quantum level spacing that is generically $\mathcal{O}(\hbar\sqrt{\kappa/\mu})$. For $\dot{\lambda}\ll\sqrt{\kappa/\mu}$, therefore, the classically inevitable failure of adiabaticity does \emph{not} occur quantum mechanically. Tunneling through the narrow peak of the barrier, at energies just below the barrier height, is easy enough that the classically singular nature of the barrier top is quantum mechanically regularized.

\subsection{The lattice of avoided crossings}\label{app:lattice_avoided_crossings}
From the simple Bohr-Sommerfeld quantization condition (\ref{oldBS}) we can see that
\begin{align}
    \tilde{S}_\mathcal{A}(E_{mn},\lambda_{mn})&\stackrel{!}{=}(m+1/2)\pi\hbar \\ \nonumber
    &=\tilde{S}_\mathcal{A}(E_{00},\lambda_{00})+\partial_E\tilde{S}_\mathcal{A} (E_{mn} - E_{00}) \\\nonumber&\hspace{0.37cm}+ \partial_\lambda\tilde{S}_\mathcal{A} (\lambda_{mn} - \lambda_{00})
\end{align}
up to corrections of higher order than first in $(E_{mn}-E_{00})$ and $(\lambda_{mn}-\lambda_{00})$, when the partial derivatives of $\tilde{S}_\mathcal{A}(E,\lambda)$ are evaluated at $(E_{00},\lambda_{00})$. A similar expansion holds for $\tilde{S}_\mathcal{C}$, with $m\to n$, so that together we can infer
\begin{align}
    \left(\begin{matrix} m\pi\hbar \\ n\pi\hbar\end{matrix}\right) = \left(\begin{matrix} \partial_E\tilde{S}_\mathcal{A} & \partial_\lambda\tilde{S}_\mathcal{A}\\ \partial_E\tilde{S}_\mathcal{C} & \partial_\lambda\tilde{S}_\mathcal{C} \end{matrix}\right)\left(\begin{matrix} E_{mn}-E_{00}\\ \lambda_{mn}-\lambda_{00} \end{matrix}\right)\;,
\end{align}
up to corrections that will self-consistently be of order $\hbar^2$ as long as $m,n=\mathcal{O}(\hbar^0)$. Inverting the 2x2 matrix yields (\ref{Elam2}) from our text.

\subsection{The Hamiltonian in the subspace of two energy levels that nearly cross}\label{app:hamiltonian}
To derive Eqn.~(\ref{LZmodel}) of our text, we consider $e^{-T_b/\hbar}$ to be small, and expand in it. The Bohr-Sommerfeld rule (\ref{GenBS}) can be rewritten as
\begin{align}
\label{GenBS1}
&\cos\left(\frac{\tilde{S}_A}{\hbar}\right)\cos\left(\frac{\tilde{S}_C}{\hbar}\right) \nonumber\\&= \sin\left(\frac{\tilde{S}_A}{\hbar}\right)\cos\left(\frac{\tilde{S}_C}{\hbar}\right)\left[\frac{\sqrt{e^{-\frac{2T_b}{\hbar}}+1}-1}{\sqrt{e^{-\frac{2T_b}{\hbar}}+1}+1}\right]
\end{align}
and in the limit $e^{-T_b/\hbar} \ll 1$, the term in the square brackets can be expanded: 
\begin{equation}
    \left[\frac{\sqrt{e^{-\frac{2T_b}{\hbar}}+1}-1}{\sqrt{e^{-\frac{2T_b}{\hbar}}+1}+1}\right]\longrightarrow \frac{e^{-\frac{2T_b}{\hbar}}}{4} + \mathcal{O}\left(e^{-\frac{4T_b}{\hbar}}\right)
\end{equation}
so we get
\begin{equation}
\label{GenBS2}
\cos\left(\frac{\tilde{S}_A}{\hbar}\right)\cos\left(\frac{\tilde{S}_C}{\hbar}\right) = \frac{e^{-\frac{2T_b}{\hbar}}}{4}\sin\left(\frac{\tilde{S}_A}{\hbar}\right)\cos\left(\frac{\tilde{S}_C}{\hbar}\right)
\end{equation}
 up to order $e^{-2T_b/\hbar}$. This is the Bohr-Sommerfeld rule one gets by using Airy-type connection formulas at the inner turning points instead of connecting the solutions to the left and right of the barrier with the connection formula based on parabolic cylinder functions as previously mentioned since the tunneling correction $\Phi$ in $\tilde{S}_\mathcal{A,C}$ is small for energies far below the top of the barrier. At zeroth order, (\ref{GenBS2}) reduces to
\begin{equation}
\label{oldBS}
     \cos\Big(\frac{\tilde{S}_\mathcal{A}}{\hbar}\Big)\cos\Big(\frac{\tilde{S}_\mathcal{C}}{\hbar}\Big)\stackrel{!}{=}0 
\end{equation}

implying, as mentioned above, the original Bohr-Sommerfeld quantization rules \emph{either} $\tilde{S}_\mathcal{A}(E_0,\lambda)\stackrel{!}{=}(m+1/2)\pi\hbar$ \emph{or} $\tilde{S}_\mathcal{C}(E_0,\lambda)\stackrel{!}{=}(n+1/2)\pi\hbar$ for integer $m,n$. These conditions define the zeroth-order semi-classical energy levels $E_0\to E^\mathcal{A}_m(\lambda)$ and $E_0\to E^\mathcal{C}_n(\lambda)$, respectively. 

As noted in the text, there is a lattice of points $(E_{mn},\lambda_{mn})$ in the $(E,\lambda)$ plane at which the zeroth-order semi-classical energy levels cross, $E^\mathcal{A}_m(\lambda_{mn})=E^\mathcal{C}_n(\lambda_{mn})=:E_{mn}$ when $E^{A}_m$ and $E^{C}_n$ are computed from (\ref{oldBS}), $e^{-T_b/\hbar}$ being neglected.

To see the non-perturbative minimal gap between these nearly crossing levels, then,  we use the expansion of the full modified Bohr-Sommerfeld condition (\ref{GenBS}) to order $e^{-2T_b(E_{mn},\lambda_{mn})/\hbar}$, which is (\ref{GenBS2}). Then, we look at $\lambda = \lambda_{mn}+e^{-T_b(E_{mn},\lambda_{mn})/\hbar} \Delta\lambda$ and we assume $E=E_{mn}+e^{-T_b(E_{mn},\lambda_{mn})/\hbar} \Delta E$. This results in the approximations
\begin{align}
\label{approxS}
    &\tilde{S}_{\mathcal{A},\mathcal{C}}\left(E_{m,n}+\varepsilon\Delta E,\lambda_{mn}+\varepsilon\Delta\lambda\right)\nonumber\\&\approx  \tilde{S}_{\mathcal{A},\mathcal{C}}\left(E_{mn},\lambda_{mn}\right) + \varepsilon\left\{\Delta E \frac{\partial\tilde{S}_{\mathcal{A},\mathcal{C}}}{\partial E} + \Delta\lambda\frac{\partial\tilde{S}_{\mathcal{A},\mathcal{C}}}{\partial\lambda}\right\}\Big|_{E_{mn},\lambda_{mn}}
\end{align}
Using the facts that $\tilde{S}_\mathcal{A}(E_{mn},\lambda_{mn})=(m+1/2)\pi\hbar$ and $\tilde{S}_\mathcal{C}(E_{mn},\lambda_{mn})=(n+1/2)\pi\hbar$ and inserting the expressions (\ref{approxS}) into (\ref{GenBS2}) this yields
\begin{align}\label{eqa1}
\Big(\Delta E\partial_E\tilde{S}_\mathcal{A} + \Delta\lambda\partial_\lambda\tilde{S}_\mathcal{A}\Big)\Big(\Delta E\partial_E\tilde{S}_\mathcal{C} + \Delta\lambda\partial_\lambda\tilde{S}_\mathcal{C}\Big)=\frac{\hbar^2}{4}\;,
\end{align}
where $T_b$ and $\partial_{E,\lambda}\tilde{S}_\mathcal{A,C}$ are all to be evaluated at $(E,\lambda)=(E_{mn},\lambda_{mn})$, and we omit corrections of higher order in $e^{-T_b/\hbar}$. Solving (\ref{eqa1}) as a quadratic equation for $\Delta E$, we obtain
\begin{align}\label{eqa2}
\Delta E = &-\frac{\Delta\lambda}{2}\left(\frac{\partial_\lambda\tilde{S}_\mathcal{A}}{\partial_E\tilde{S}_\mathcal{A}}+\frac{\partial_\lambda\tilde{S}_\mathcal{C}}{\partial_E\tilde{S}_\mathcal{C}}\right)\nonumber\\
&\pm \frac{\sqrt{\hbar^2\partial_E\tilde{S}_\mathcal{A}\partial_E\tilde{S}_\mathcal{C} + \Delta\lambda^2[\tilde{S}_\mathcal{A},\tilde{S}_\mathcal{C}]^2}}{2\partial_E\tilde{S}_\mathcal{A}\partial_E\tilde{S}_\mathcal{C}}\;,
\end{align}
where the definition of the Poisson-like bracket (\ref{poisson}) from our main text was used. The expressions (\ref{eqa2}) can be recognized as the energy eigenvalues of the $\Delta\lambda$-dependent two-state Hamiltonian
\begin{align}\label{hhh}
\hat{h}(\Delta\lambda) =& -\frac{\Delta\lambda}{2}\left(\frac{\partial_\lambda\tilde{S}_\mathcal{A}}{\partial_E\tilde{S}_\mathcal{A}}+\frac{\partial_\lambda\tilde{S}_\mathcal{C}}{\partial_E\tilde{S}_\mathcal{C}}\right)\nonumber\\
&+\frac{\Delta\lambda}{2}\left(\frac{\partial_\lambda\tilde{S}_\mathcal{A}}{\partial_E\tilde{S}_\mathcal{A}}-\frac{\partial_\lambda\tilde{S}_\mathcal{C}}{\partial_E\tilde{S}_\mathcal{C}}\right)\hat{\sigma}_z
+\frac{\hbar \hat{\sigma}_x}{2\sqrt{\partial_E\tilde{S}_\mathcal{A}\partial_E\tilde{S}_\mathcal{C}}}\;
\end{align}
Since the $\mathcal{O}(e^{-T_b/\hbar})^0$ energies $E^{A}_m$ and $E^\mathcal{C}_n$ are defined by the simple Bohr-Sommerfeld conditions (\ref{oldBS}), by differentiating with respect to $\lambda$ we find
\begin{align}\label{ddd}
    \frac{d}{d\lambda}\tilde{S}_{A} = \frac{d}{d\lambda}(m+1/2)\pi\hbar &=0\nonumber\\
    \Longrightarrow\;
    \frac{d}{d\lambda}E^\mathcal{A}_m  &= - \frac{\partial_\lambda \tilde{S}_\mathcal{A}}{\partial_E{\tilde{S}_\mathcal{A}}}\;,
\end{align}
and similarly for $\tilde{S}_\mathcal{C}$ and $E^\mathcal{C}_n$. Using this result (\ref{ddd}) in (\ref{hhh}), restoring the factors of $e^{-T_b/\hbar}$ and expressing $\lambda-\lambda_{mn}$ as $\dot{\lambda}(t-t_{mn})$, we realize the time-dependent two-state Hamiltonian for each avoided crossing as given by (\ref{LZmodel}) in our text. And at the same time we can use the fact from (\ref{ddd}) that
\begin{equation}
    \partial_E\tilde{S}_\mathcal{A}\partial_E\tilde{S}_\mathcal{C}\Big|\frac{dE^\mathcal{A}_m}{d\lambda}-\frac{dE^\mathcal{C}_n}{d\lambda}\Big|=\Big|[\tilde{S}_\mathcal{A},\tilde{S}_\mathcal{C}]\Big|
\end{equation}
to derive $e^{-2\pi\gamma^2/\nu^2}=P(E,\lambda)$ as stated in (\ref{PElam}).

\subsection{The probability lattice}\label{app:probability_lattice}
We have thus found the probability $P(E_{mn},\lambda_{mn})$ of diabatic Landau-Zener transitions at each $(m,n)$ avoided crossing, in terms of $E_{mn}$ and $\lambda_{mn}$ (including $\dot{\lambda}(t_{mn})$, considered as a function of $\lambda$ evaluated at $\lambda_{mn}$). To obtain our text's centrally important equation (\ref{Pmn}) for the probabilities $P_{mn}$ in an $\mathcal{O}(\hbar^{-1})$-sized portion of the lattice of crossings, we identify $(m,n)=(0,0)$ with one crossing near the quantum separatrix, and evaluate $P(E_{mn},\lambda_{mn})$ for $m,n$ of order $\hbar^0$.

Since (\ref{Elam2}) tells us that $E_{mn}=E_{00}+\mathcal{O}(\hbar)$ and $\lambda_{mn}=\lambda_{00}+\mathcal{O}(\hbar)$ for $m,n = \mathcal{O}(\hbar^0)$, we can conclude that in the expression (\ref{PElam}) for $P_{mn}=P(E_{mn},\lambda_{mn})$ we can write 
\begin{align}
    \dot{\lambda}(t_{mn})&=\dot{\lambda}(t_{00}) + \mathcal{O}(\hbar)\;\mathrm{and}\nonumber\\
    [\tilde{S}_\mathcal{A},\tilde{S}_\mathcal{C}]\Big|_{E_{mn},\lambda_{mn}} &=[\tilde{S}_\mathcal{A},\tilde{S}_\mathcal{C}]\Big|_{E_{00},\lambda_{00}}+ \mathcal{O}(\hbar)\;.
\end{align}
For the factor $e^{-2T_b/\hbar}$ in the exponent of $P(E_{mn},\lambda_{mn})$, however, the explict factor of $1/\hbar$ means that we must write
\begin{widetext}
\begin{align}
    e^{-\frac{2}{\hbar}T_b(E_{mn},\lambda_{mn})} &= e^{-\frac{2}{\hbar}T_b(E_{00},\lambda_{00})}\exp\Big[-2\frac{(E_{mn}-E_{00})\partial_E T_b+(\lambda_{mn}-\lambda_{00})\partial_\lambda T_b}{\hbar}\Big]\times[1+\mathcal{O}(\hbar)]\nonumber\\
    &\equiv e^{-\frac{2}{\hbar}T_b(E_{00},\lambda_{00})}\exp\Big[-2\frac{m[T_b,\tilde{S}_\mathcal{C}]+n[T_b,\tilde{S}_\mathcal{A}]}{[\tilde{S}_\mathcal{A},\tilde{S}_\mathcal{C}]}\Big]\times[1+\mathcal{O}(\hbar)]
\end{align}
\end{widetext}
Dropping the explicit $+\mathcal{O}(\hbar)$ thus recovers (\ref{Pmn}) for $P_{mn}$. 

It is interesting to see how the non-perturbative $1/\hbar$ factor in the exponent of the quantum tunneling amplitude $e^{-T_b/\hbar}$ has conspired with energy quantization in steps of $\mathcal{O}(\hbar)$ to make the crucial $X$ and $Y$ parameters in $P_{mn}$ independent of $\hbar$. The result is that although the quantum separatrix is qualitatively non-classical in nature, being defined by a coherent random walk through quantized energy levels with non-zero amplitude for tunneling, and although its location in the $(E,\lambda)$ plane depends non-classically on the sweep rate $\dot{\lambda}$ and $\hbar$, yet the width and slope of the quantum separatrix are determined in the semi-classical limit by quantities $X$ and $Y$ that are entirely classical, in the sense that they are composed of derivatives of action integrals with respect to $E$ and $\lambda$, without involving $\hbar$.

\section{Lattice geometry and rates of change of the subspace dimensionalities}
\subsection{lattice geometry}\label{app:lattice_geometry}
\begin{figure}
    \centering
    \includegraphics[scale=0.25]{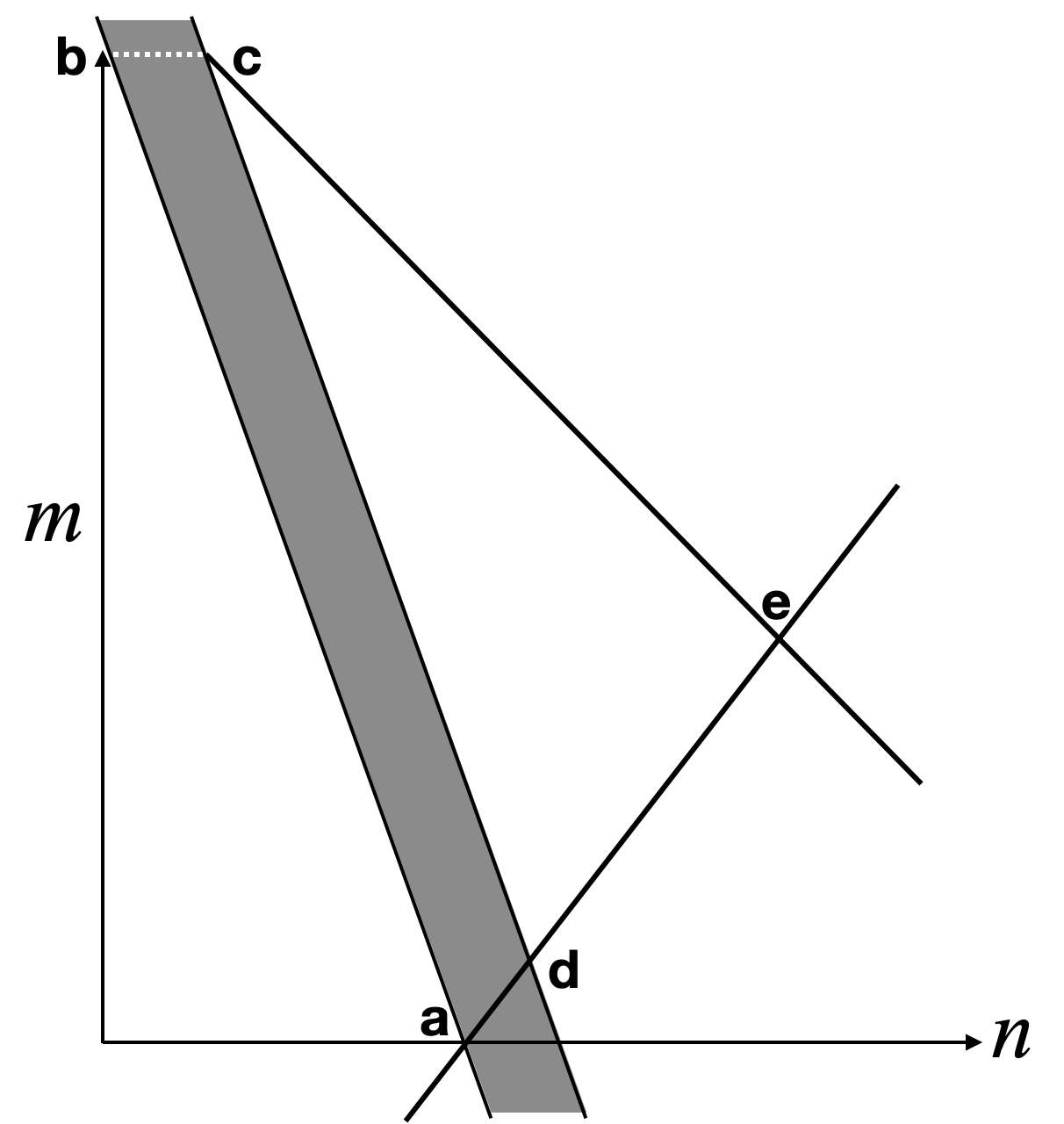}
    \caption{Geometry of the lattice of avoided crossings in the $(n,m)$ plane. The coordinate system is rotated by $45^\circ$ compared to Figure \ref{fig:TheoremFig}. The number of adiabatic levels above the separatrix zone $K$, by the time the entire initial ensemble has entered the separatrix zone, in eq. (\ref{KN_geom}) is given by the sum $\Delta m + \Delta n$ of the vector connecting the points $\vec{e}$ and $\vec{d}$ across the separatrix zone along the line of constant $\lambda$ which is $\sqrt{2}$ times the magnitude of the projection of the vector $\vec{e}-\vec{d}$ in the direction $(1,1)/\sqrt{2}$. The points $a$ to $e$ will be used as an aid to geometrically derive this vector. The thick gray bar is the separatrix zone which contains D levels.}
    \label{fig:geometry}
\end{figure}
When the $(E,\lambda)$ plane is viewed in continuous $(n,m)$ coordinates as in Fig.~\ref{LatticeFig}b), (\ref{Pmn}) tells us that the separatix $(E,\lambda)=(E_s(\lambda),\lambda)$ is parallel to the vector $(n,m)=(X,-Y)$. From (\ref{Elam2}) it follows, also, that lines of constant $\lambda$ in the $(n,m)$ plane are parallel to the vector $(\partial_E \tilde{S}_\mathcal{C},\partial_E\tilde{S}_\mathcal{A})$. With the $45^\circ$-rotated axes of Figs.~\ref{LatticeFig}b) and \ref{fig:TheoremFig}, these $\lambda$-lines must always be closer to vertical than horizontal, because both components of $(\partial_E \tilde{S}_\mathcal{C},\partial_E\tilde{S}_\mathcal{A})$ are positive. Time thus always runs roughly rightwards in any figures like Fig.~\ref{fig:TheoremFig}.\newline
In the $(n,m)$ representation with axes rotated to face up, as in Fig.~\ref{fig:TheoremFig}, the adiabatic Bohr-Sommerfeld levels are by construction all lines parallel to the $m$ and $n$ axes, \textit{i.e.} at $\pm 45^\circ$. The separatrix width in levels $D$ is thus $\sqrt{2}$ times the width of the separatrix zone in the vertical direction $(n,m)=(1,1)/\sqrt{2}$ as indicated in Fig.~\ref{fig:TheoremFig}. 
From similar geometric considerations it follows that the number of $|E^\mathcal{C}_n\rangle$ levels below the separatrix zone, into which the initial ensemble might have evolved by the time the entire initial ensemble has entered the separatrix zone, is 
\begin{equation}\label{Leq1}
N = M\times \frac{X}{Y}+\delta N
\end{equation}
where $|\delta N|<1$ comes from discretization ($N$ must be an integer). So also is the number $K$ of zig-zagging adiabatic levels above the separatrix zone, into which the initial ensemble might have evolved by this time, given by 
\begin{align}\label{Keq1}
K &= (M-kD)\times\Big( 1- \frac{X}{Y}\Big) + \delta K\nonumber\\
k &= \frac{\partial_E\tilde{S}_\mathcal{A}}{\partial_E\tilde{S}_\mathcal{A}+\partial_E\tilde{S}_\mathcal{C}}\;,
\end{align}
where $\delta K$ again comes discretization, while the $-k D$ comes from that fact that the lines of constant $\lambda$ are not parallel to the $n$ axis. In order to simplify the geometrical considerations, we rotate these axes by $45^\circ$ compared to Fig.~\ref{fig:TheoremFig} as can be seen in Fig.~\ref{fig:geometry}. 
In the following, (\ref{Keq1}) will be derived geometrically by considering the vectors $\vec{a}-\vec{e}$ (see Fig. \ref{fig:geometry}) and parametrizing the lines connecting them in the $(n,m)$ coordinates:
\begin{align}
\label{app:geom:points:a}
    &\Vec{a} = (N,0) = \left(\frac{MX}{Y},0\right)\\\label{app:geom:points:b}
    &\Vec{b} = (0,M)\\\label{app:geom:points:ab}
    &\overline{ab} = (N,0) + r(X,-Y)\\\label{app:geom:points:bc}
    &\overline{bc} = (0,M) + s(1,0)\\\label{app:geom:points:d}
    &\Vec{d} = \left(N,0\right) + \frac{D}{\partial_E\tilde{S}_\mathcal{C}+\partial_E\tilde{S}_\mathcal{A}}\left(\partial_E\tilde{S}_\mathcal{C},\partial_E\tilde{S}_\mathcal{A}\right)\\\label{app:geom:points:cd}
    &\overline{cd} = (N,0) +  t(X,-Y)+\frac{D}{\partial_E\tilde{S}_\mathcal{C}+\partial_E\tilde{S}_\mathcal{A}}\left(\partial_E\tilde{S}_\mathcal{C},\partial_E\tilde{S}_\mathcal{A}\right)\\\label{app:geom:points:ae}
    &\overline{ae} = (N,0) + u\left(\partial_E\tilde{S}_C,\partial_E\tilde{S}_A\right)\\ \label{app:geom:points:ce}
    &\overline{ce} = \Vec{c} + v(1,-1)
\end{align}
Here, the fact that lines of constant $\lambda$ in the $(n,m)$ plane are proportional to the vector $\left(\partial_E\tilde{S}_C,\partial_E\tilde{S}_A\right)$ has been used again. It may be noted that the width of the separatrix zone in levels D is defined by the sum $\Delta m + \Delta n$ of the vector connecting one side of the separatrix zone with the other along the line of constant $\lambda$. This vector is given by the vector connecting $\vec{d}$ and $\vec{a}$ in Fig.~\ref{fig:geometry}. The the number of levels $D$ within the separatrix zone is thus given by $D = \left(\vec{d}-\vec{a}\right)\cdot\left(1,1\right)$. 
\newline The fact that $\Vec{c}$ is on both of the connecting lines $\overline{bc}$ and $\overline{cd}$ (see (\ref{app:geom:points:bc}), (\ref{app:geom:points:cd})) can be used to determine the parameters t and s and ultimately the point $\vec{c}$:
\begin{align}
    &\Vec{c} = (s,M) = \left(N + tX,-tY\right) + D\frac{\left(\partial_E\tilde{S}_\mathcal{C},\partial_E\tilde{S}_\mathcal{A}\right)}{\partial_E\tilde{S}_\mathcal{A}+\partial_E\tilde{S}_\mathcal{C}}\nonumber\\
    &\Longrightarrow t = \frac{1}{Y}\left(D\frac{\partial_E\tilde{S}_\mathcal{A}}{\partial_E\tilde{S}_\mathcal{A}+\partial_E\tilde{S}_\mathcal{C}}-M\right)\nonumber\\
    &\Longrightarrow s = N + \frac{D}{\partial_E\tilde{S}_\mathcal{A}+\partial_E\tilde{S}_\mathcal{C}}\left(\partial_E\tilde{S}_\mathcal{C}+\frac{X}{Y}\partial_E\tilde{S}_\mathcal{A}\right)-M\frac{X}{Y}\nonumber\\
    &\Longrightarrow \Vec{c} = \left(\frac{D}{Y}\frac{X\partial_E\tilde{S}_\mathcal{A}+Y\partial_E\tilde{S}_\mathcal{C}}{\partial_E\tilde{S}_\mathcal{A}+\partial_E\tilde{S}_\mathcal{C}},M\right)
\end{align}
Analogously, we can determine the $(n,m)$ coordinates of the point $\Vec{e}$ by using the fact that it lies on the connecting lines $\overline{ce}$ and $\overline{ae}$:
\begin{align}
\Vec{e} &= \left(\frac{D}{Y}\frac{X\partial_E\tilde{S}_\mathcal{A}+Y\partial_E\tilde{S}_\mathcal{C}}{\partial_E\tilde{S}_\mathcal{A}+\partial_E\tilde{S}_\mathcal{C}},M\right)+v\left(1,-1\right)\nonumber \\&= \left(N,0\right)+u\left(\partial_E\tilde{S}_\mathcal{C},\partial_E\tilde{S}_\mathcal{A}\right)
\end{align}
Adding the two equations for the $m,n$ components results in 
\begin{align}
    &M+ \frac{D}{Y}\frac{X\partial_E\tilde{S}_\mathcal{A}+Y\partial_E\tilde{S}_\mathcal{C}}{\partial_E\tilde{S}_\mathcal{A}+\partial_E\tilde{S}_\mathcal{C}} = N+u\left(\partial_E\tilde{S}_\mathcal{C} + \partial_E\tilde{S}_\mathcal{A}\right)\nonumber\\
    &\Longrightarrow u = \frac{1}{\partial_E\tilde{S}_\mathcal{A}+\partial_E\tilde{S}_\mathcal{C}}\left\{M\left(1-\frac{X}{Y}\right)+\frac{D}{Y}\frac{X\partial_E\tilde{S}_\mathcal{A}+Y\partial_E\tilde{S}_\mathcal{C}}{\partial_E\tilde{S}_\mathcal{A}+\partial_E\tilde{S}_\mathcal{C}}\right\}
\end{align}
which can be inserted into (\ref{app:geom:points:ae}) to obtain the point $\vec{e}$. With this, we can finally compute the number of adiabatic levels above the separatrix zone $K$ as the sum $\Delta m + \Delta n$ of the vector connecting the points $\vec{e}$ and $\vec{d}$:
\begin{align}
    K =& \left(\Vec{e}-\Vec{d}\right)\cdot\left(1,1\right)\\
    =&\left(u-\frac{D}{\partial_E\tilde{S}_\mathcal{A}+\partial_E\tilde{S}_\mathcal{C}}\right)\left(\partial_E\tilde{S}_\mathcal{A}+\partial_E\tilde{S}_\mathcal{C}\right)\\
    =&M\left(1-\frac{X}{Y}\right)+D\frac{\left(\frac{X}{Y}-1\right)\partial_E\tilde{S}_\mathcal{A}}{\partial_E\tilde{S}_\mathcal{A}+\partial_E\tilde{S}_\mathcal{C}}\\
    =&\left(M-kD\right)\left(1-\frac{X}{Y}\right)
\end{align}
with $k$ given by eq. (\ref{Keq1}). 
\subsection{rates of change of the subspace dimensionalities}\label{app:growthrates}
Since the quantum separatrix runs parallel to the vector $(-Y,X)$ in the $(m,n)$ plane, we can parametrize $(m_s,n_s)$ on the separatrix:
\begin{align}
    &m_s = m0 + \mu_s\\
    &n_s = n0 + \nu_s
\end{align}
with $\mu_s = -Ys$ and $\nu_s = Xs$. The average rates of change of the dimensionalities $D_\mathcal{A,B,C}$ of the respective subspaces are thus given by
\begin{align}
    &\frac{dD_\mathcal{A}}{d\lambda} = \frac{d\mu_s}{d\lambda} = -Y\frac{ds}{d\lambda}\\
    &\frac{dD_\mathcal{C}}{d\lambda} = \frac{d\nu_s}{d\lambda}= X\frac{ds}{d\lambda}\\
    &\frac{dD_\mathcal{B}}{d\lambda} = -\left(\frac{dD_\mathcal{A}}{d\lambda}+\frac{dD_\mathcal{C}}{d\lambda}\right)
\end{align}
Equation (\ref{Elam2}) then gives the relation between $\lambda$ and the separatrix parameter s:
\begin{align}
    \lambda(s) &= \lambda_{00} + \pi\hbar\frac{\nu_s\partial_E \tilde{S}_\mathcal{A} - \mu_s\partial_E\tilde{S}_\mathcal{C}}{[\tilde{S}_\mathcal{A},\tilde{S}_\mathcal{C}]} \nonumber\\
    &= \lambda_{00} + \pi\hbar s\frac{X\partial_E \tilde{S}_\mathcal{A} + Y\partial_E\tilde{S}_\mathcal{C}}{[\tilde{S}_\mathcal{A},\tilde{S}_\mathcal{C}]} 
\end{align}
From this, one can easily compute
\begin{equation}\label{lambdadiff}
    \frac{1}{\Gamma} = \frac{d\lambda}{ds} = \pi\hbar\frac{\left(X\partial_E\tilde{S}_\mathcal{A}+Y\partial_E\tilde{S}_\mathcal{C}\right)}{\left[\tilde{S}_\mathcal{A},\tilde{S}_\mathcal{C}\right]}
\end{equation}
\section*{References}

\bibliography{bibliography}
\end{document}